\documentclass[preprint,12pt]{elsarticle}

\usepackage{graphicx}

\usepackage{amssymb}

\journal{JQSRT}

\begin{document}

\begin{frontmatter}



\title{\emph{T}-matrix calculation via discrete-dipole approximation, point matching and exploiting symmetry}

\author{Vincent L. Y. Loke}
 \ead{loke@physics.uq.edu.au}
\author{Timo A. Nieminen}
\author{N. R. Heckenberg}
\author{Halina Rubinsztein-Dunlop}


\address{Centre for Biophotonics and Laser Science, School of Physical Sciences, The University of Queensland, Brisbane, Queensland 4072, Australia.}

\begin{abstract}
We present a method of incorporating the discrete dipole approximation (DDA) method with the point matching method to formulate the \emph{T}-matrix for modeling arbitrarily shaped micro-sized objects. The \emph{T}-matrix elements are calculated using point matching between fields calculated using vector spherical wave functions and DDA. When applied to microrotors, their discrete rotational and mirror symmetries can be exploited to reduce memory usage and calculation time by orders of magnitude; a number of optimization methods can be employed based on the knowledge of the relationship between the azimuthal mode and phase at each discrete rotational point, and mode redundancy from Floquet's theorem. A `reduced-mode' \emph{T}-matrix can also be calculated if the illumination conditions are known.
\end{abstract}

\begin{keyword}
DDA \sep T-matrix \sep discrete rotational symmetry \sep point matching
\end{keyword}

\end{frontmatter}




\section{\label{Intro}Introduction}
Optical tweezers~\cite{Ashkin1986} can be used to exert forces and torques and thus drive micromachines~\cite{Grier2003,Nieminen2008a}. This opens up a new field of micro engineering, whose potential has yet to be fully realized. To aid in designing micromachines, we employ a number of modelling methods ranging from the generalized Lorenz-Mie theory (GLMT)~\cite{Nieminen2007a}, the FDFD/\emph{T}-matrix hydrid method~\cite{Loke2007a} to the discrete dipole approximation (DDA)~\cite{Purcell1973,Draine1994}. We use the abovementioned methods to formulate the \emph{T}-matrix because repeated calculations with different illumination conditions are required. The extended boundary condition method (EBCM)~\cite{Waterman1965,Waterman1971} is most commonly used to calculate the \emph{T}-matrix; however, numerical problems are encountered with high aspect ratio structures~\cite{Mishchenko2002}. 

\begin{figure}[ht]
\centerline{%
(a)\includegraphics[height=5cm]{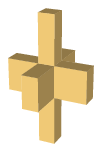}
\hfill
(b)\includegraphics[height=5cm]{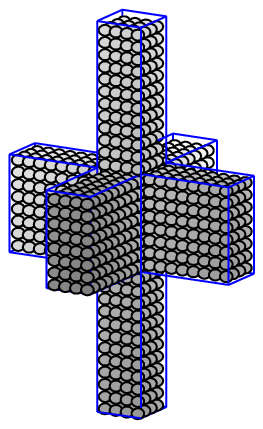}
\hfill
(c)\includegraphics[height=5cm]{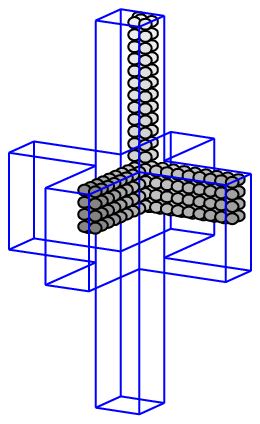}}
\caption{(a)~An optically-driven microrotor. (b)~Discrete dipole representation of the rotor.
(c)~Discrete rotational and mirror symmetries allow the modelling of only the repeated segment,
with a major reduction in required memory and time.}
\label{fig:cross}
\end{figure}

Given the structure of typical optical micromachines~\cite{Asavei2008a}, we have found the DDA method most suitable. In DDA, the scattering object is represented as a collection of dipole scatterers, and the total scattering problem, including the coupling between the dipoles, is solved. DDA is well-suited to modelling optical micromachines. Firstly, only
the volume of the actual particle needs to be discretized, while both the particle and surrounding medium in a volume enclosing the particle are
discretized in other general methods such as the finite-difference time-domain method (FDTD) and finite element methods (FEM). Considering that structures as
shown in figure \ref{fig:cross}a are not unusual, where the particle occupies only a relatively small fraction of the nearby volume, this can mean a considerable
saving in required memory and time. Secondly, DDA performs well for relatively low contrast scatterers, which is typical of most optical micromachines so far, usually constructed from a polymer material~\cite{Asavei2009a} and deployed in a dielectric liquid. Thirdly, it is relatively simple to obtain the \textit{T}-matrix
via DDA if repeated calculations are desired~\cite{Mackowski2002}. Finally, it is possible to exploit discrete rotational symmetry of a particle to reduce
the computational resources, including both time and memory, by orders of magnitude. This last factor is important, since optical micromachines
are often large in overall dimension compared to the wavelength (while having wavelength scale features forcing the use of electromagnetic
theory rather than geometric optics), and the available resources can place such devices beyond practical calculation~\cite{Collett2003}. Figure~\ref{fig:cross} shows a typical case.

By itself, the DDA method does not give us the \emph{T}-matrix. One method of formulating the \emph{T}-matrix via DDA~\cite{Mackowski2002} is to transform the field contributions of each dipole to aggregate their contributions at a common origin. We embarked on a different method --- incorporated the DDA with the point matching method; both the near and far-field point matching methods are discussed in this paper but only the results of the former are presented. We will also discuss the optimization methods employed.
\section{DDA revisited}
With DDA, the microrotor in figure \ref{fig:cross}a can be represented by a set of dipoles in a cubic lattice, as in figure \ref{fig:cross}b, fashioned to approximate the target structure. The lattice spacing has to be sufficiently small, not only to accurately represent the structure --- stair-casing is becomes a problem for curved surfaces --- but small enough compared to the wavelength of the incident light; the criterion is specified in \cite{Draine1994}.

We first have to calculate the incident field, $\mathbf{E}_{inc,j}$, at each dipole and the interaction matrix~\cite{Draine1994}
\begin{equation}
\label{eqn:A_offdiag}  
\mathbf{A}_{jk}=\frac{\exp(\mathrm{i}kr_{jk})}{r_{jk}}\left[k^2(\hat{r}_{jk}\hat{r}_{jk}-\mathbf{1}_3)
+ \frac{\mathrm{i}kr_{jk}-1}{r^2_{jk}}(3\hat{r}_{jk}\hat{r}_{jk}-\mathbf{1}_3)\right],
\; j\neq k,
\end{equation}
where $r_{jk}$ is the distance from points $r_j$ to $r_k$, $\hat{r}_{jk}$ is the unit vector from in the direction from points $r_j$ to $r_k$, and
\begin{equation}
\label{eqn:A_ondiag}  
\mathbf{A}_{jj} = \alpha^{-1}_j
\end{equation}
where $\alpha_j$ is the polarizability~\cite{Draine1993} of each dipole. Equation (\ref{eqn:A_offdiag}) represents the coupling between the dipoles and (\ref{eqn:A_ondiag}) allows the system of equations to be written in the compact form
\begin{equation}
\label{eqn:AP_E}
\sum^{N}_{k=1}\mathbf{A}_{jk}\mathbf{P}_j=\mathbf{E}_{inc,j}.
\end{equation}
since the dipole moment due to the total field at each dipole is $\mathbf{P}_j=\alpha_j\mathbf{E}_j$. Because $\mathbf{A}_{jk}$ is a square matrix, the dipole moments can rapidly solved using the generalized minimum residual method~\cite{Saad1986}.
\section{Optimizing the Discrete Dipole Approximation interaction matrix}
\label{DDA_rotsym}
\begin{figure}[ht!]
\centerline{%
(a)~\includegraphics[height=8cm]{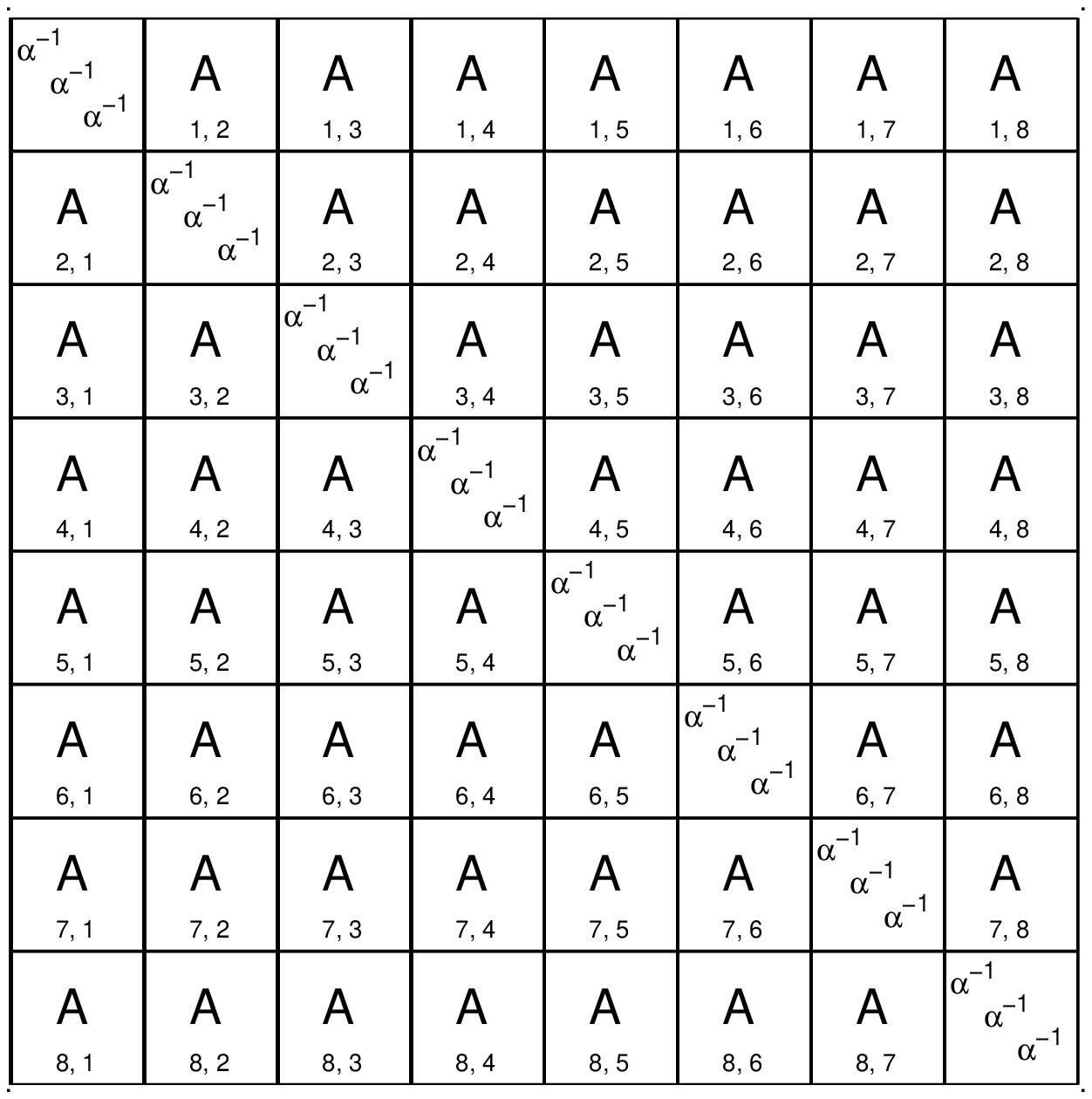}
\hfill\hfill\hfill
(b)~\includegraphics[height=2.4cm]{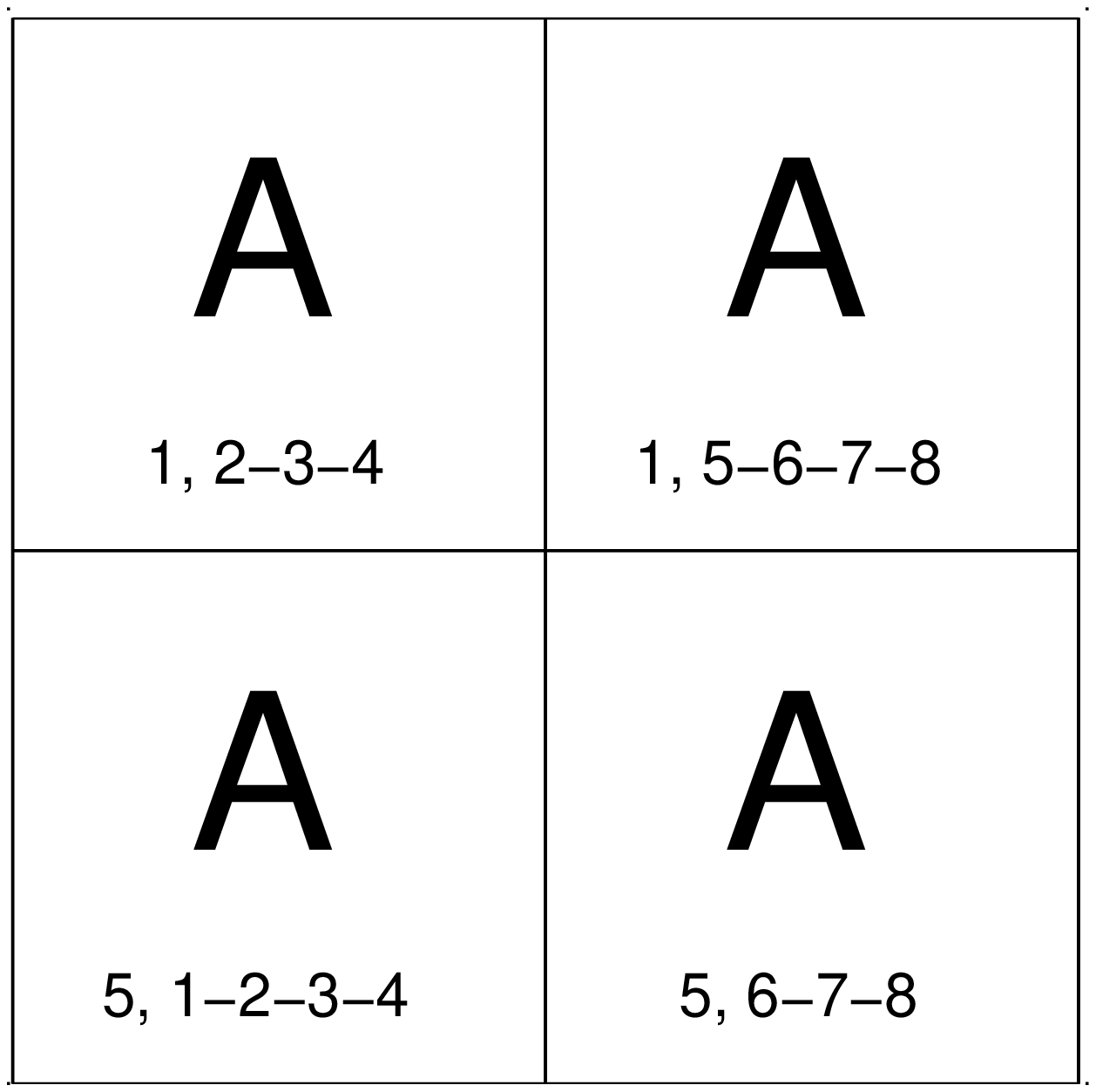}}
\caption{(a)~Full interaction matrix. (b)~Compressed interaction matrix.}
\label{fig_interaction}
\end{figure}

The size of the microdevices we model may exceed 10--20 wavelengths in size, which may well require computational time in excess of several days and RAM beyond that available. To circumvent these limitations, we exploit the discrete rotational and/or mirror symmetry of a microcomponent. This is closely tied with the link between DDA and the \textit{T}-matrix method. In the \textit{T}-matrix method, the fields are represented as sums of vector spherical wavefunctions (VSWFs)~\cite{Waterman1971,Nieminen2003b}, and to use DDA to calculate a \textit{T}-matrix, we can simply calculate the scattered field (and its VSWF representation) for each possible incident single-mode VSWF field in turn. The important point is that each VSWF is characterized by a simple azimuthal dependence of $\exp(\mathrm{i}m\phi)$, where $m$ is the azimuthal mode index.
\begin{figure}[ht!]
\centerline{%
\includegraphics[height=4cm]{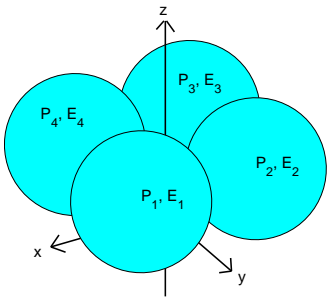}}
\caption{Dipoles in a 4-fold discrete rotationally symmetric arrangement about beam axis.}
\label{fig_rot_PE}
\end{figure}
If we consider a group of dipoles that are rotationally symmetric about the vertical axis---in the case of figure \ref{fig_rot_PE}, there is 4th-order rotational symmetry---the magnitude of the incident field will be the same, the field differing by the vector rotation (\ref{eqn_rotation}) and phase factor of $\exp(\mathrm{i}m\phi)$; only the dipole moment of one repeating unit of the total number of dipoles needs to be known. In spherical coordinates,
\begin{equation}
\label{eqn:rotsym_P}
 \mathbf{P}_q^{sph} = \mathbf{P}_1^{sph} \exp(\mathrm{i}m\phi_q).
\end{equation}
However, our implementation of the interaction matrix\cite{Draine1994}, electric field and dipole moments were in cartesian coordinates system. This requires a transformation of the dipole moment at the first segment from the cartesian to spherical coordinate system followed by a transformation back from spherical to cartesian at the rotational counterpart dipole,
\begin{equation}
\label{eqn:rotsym_P_cart}
 \mathbf{P}_q^{cart} = \mathbf{C}_q \mathbf{S} \mathbf{P}_1^{cart} \exp(\mathrm{i}m\phi_q),
\end{equation}
where 
\begin{equation}
\label{eqn:cart2sph_mat}
\mathbf{S} = \left[ \begin{array}{ccc}
\sin(\theta)\cos(\phi_1) & \sin(\theta)\sin(\phi_1) & \cos(\theta) \\
\cos(\theta)\cos(\phi_1) & \cos(\theta)\sin(\phi_1) & -\sin(\theta) \\
-\sin(\phi_1) & \cos(\phi_1) & 0 \end{array} \right]
\end{equation}
is an orthogonal matrix that transforms the $\mathbf{P}_1^{cart}$ vector into the spherical coordinate system and
\begin{equation}
\label{eqn:sph2cart_mat}
\mathbf{C}_q = \left[ \begin{array}{ccc}
\sin(\theta)\cos(\phi_q) & \sin(\theta)\sin(\phi_q) & \cos(\theta) \\
\cos(\theta)\cos(\phi_q) & \cos(\theta)\sin(\phi_q) & -\sin(\theta) \\
-\sin(\phi_q) & \cos(\phi_q) & 0 \end{array} \right]^{-1}
\end{equation}
is a transpose of $\mathbf{S}$ but at the azimuthal coordinate of the rotational counterpart dipole. The same transformations and phase corrections are applied to the $\mathbf{E}$ and $\mathbf{H}$ fields.
\begin{figure}[ht!]
\centerline{%
(a)\includegraphics[height=4cm]{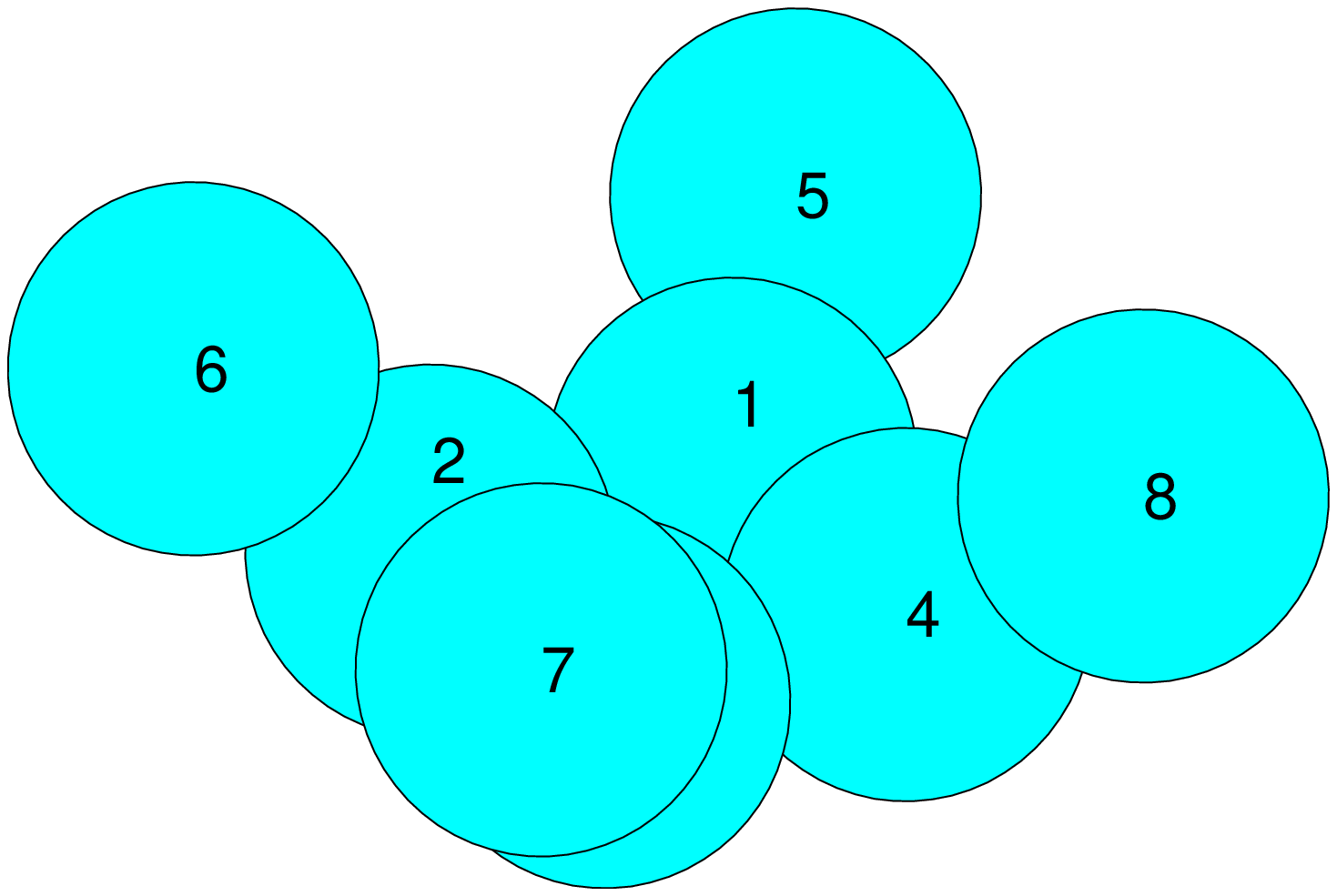}
\hfill
(b)\includegraphics[height=4cm]{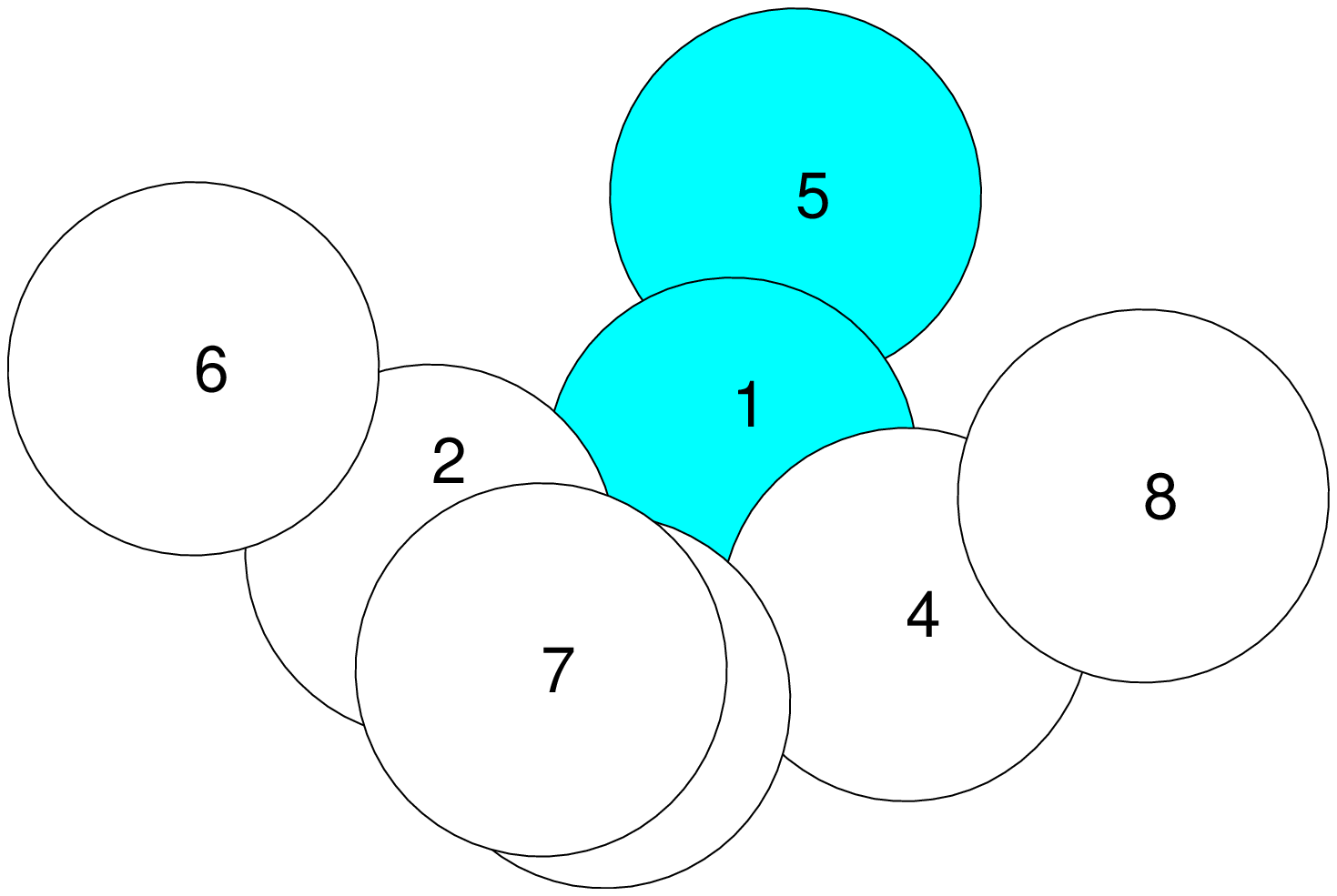}}
\caption{(a)~8-dipole example. (b)~The 2 dipoles required to completely specify all dipole moments.}
\label{fig_rot}
\end{figure}
Conventionally, the interaction matrix as defined in \cite{Draine1994} represents the coupling between each dipole with all other dipoles. Figure \ref{fig_interaction}(a) shows the interaction matrix for the example set of dipoles shown in figure~\ref{fig_rot}(a). In general the matrix will be made up of $N\times N$ cells for $N$ dipoles; each cell is a $3\times 3$ tensor. In the example, the matrix is made up of $8\times 8$ cells. A diagonal cell represents the self interaction (or the inverse of the polarizability) and an off-diagonal cell represents the coupling between different dipoles. Taking advantage of the equal amplitudes and known phase factors between
a dipole and its rotational counterparts, we can reduce the interaction matrix. Taking the example in figure \ref{fig_rot}(b), we contruct the interaction matrix as if there were only 2 dipoles but we aggregate the contribution from the appropriate dipoles. For the off-diagonal cells, the coupling between a dipole with the other dipoles including their rotational counterparts are summed as follows 
\begin{eqnarray}
\label{eqn_A_offdiag}
\mathbf{A}_{jk}&=&\sum^Q_{q=1}\frac{\exp(\mathrm{i}kr_{jk}^{(q)})}{r_{jk}^{(q)}}\nonumber\\
&&\times\left[k^2(\hat{r}_{jk}^{(q)}\hat{r}_{jk}^{(q)}-\mathbf{1}_3) + \frac{\mathrm{i}kr_{jk}^{(q)}-1}{r_{jk}^{2(q)}}(3\hat{r}_{jk}^{(q)}\hat{r}_{jk}^{(q)}-\mathbf{1}_3)\right]\nonumber\\
&&\times \; \mathbf{C}_q \mathbf{S} \exp(\mathrm{i}m\phi), \; j\neq k,
\end{eqnarray}
where $Q$ is the order of discrete rotational symmetry, $m$ is the azimuthal mode of the incident VSWF field, $q$ is the rotational segment number, the rotational angle $\phi = 2\pi q/Q$, $r_{jk}^{(q)}$ is the
distance from points $r_j$ to the rotationally symmetric points $r_k^{(q)}$, and $\hat{r}_{jk}^{(q)}$ is the unit vector from points $r_j$ to $r_k^{(q)}$. The coordinate for a given rotational symmetric point is calculated using a rotation about the z-axis,
\begin{equation}
\label{eqn_rotation}
 r_k^{(q)} = \left[ \begin{array}{ccc}
\cos(q\phi) & -\sin(q\phi) & 0 \\
\sin(q\phi) & \cos(q\phi) & 0 \\
0 & 0 & 1 \end{array} \right] r_k.
\end{equation}
For the diagonal cells, the ``self interaction'' includes the coupling between a dipole and its rotational counterparts: 
\begin{eqnarray}
\label{eqn_A_diag}
\mathbf{A}_{jj}&=&\alpha^{-1}_j + \sum^Q_{q=2}\frac{\exp(\mathrm{i}kr_{jk}^{(q)})}{r_{jk}^{(q)}}\nonumber\\
&&\times\left[k^2(\hat{r}_{jk}^{(q)}\hat{r}_{jk}^{(q)}-\mathbf{1}_3) + \frac{\mathrm{i}kr_{jk}^{(q)}-1}{r_{jk}^{2(q)}}(3\hat{r}_{jk}^{(q)}\hat{r}_{jk}^{(q)}-\mathbf{1}_3)\right]\nonumber\\
&&\times \; \mathbf{C}_q \mathbf{S} \exp(\mathrm{i}m\phi), \; j=k.
\end{eqnarray}
Figure~\ref{fig_interaction}(b) shows the interaction matrix representation for the example dipole system in figure~\ref{fig_rot}(c). The compressed interaction matrix is a factor of $Q^2$ smaller than the conventional matrix.
Having precalculated the incident fields $\mathbf{E}_{j,inc}$ at each dipole of the rotational unit, we solve for the dipole moments $\mathbf{P}_j$ for the dipoles with a reduced set of linear equations: 
\begin{equation}
\label{eqn:AP_E_quad} 
\sum^{N}_{k=1}\mathbf{A}_{quad,jk}(m)\mathbf{P}_{quad,j}=\mathbf{E}_{inc,quad,j}.
\end{equation}
Notice that the discrete rotationally optimized A-matrix is $m$ dependent; the A-matrices can be precalculated and loaded as cycle the required azimuthal modes ($m$) of the incident beam. However, there is some degree of redundancy; e.g., when $m$ is even, $\mathbf{A}(m)=\mathbf{A}(-m)$. The dipole moments of the rotational counterpart dipoles can be calculated by applying the rotational matrix \ref{eqn_rotation} and phase factor $\exp(\mathrm{i}m\phi)$.

We can exploit mirror symmetry in a similar fashion, since a VSWF possess either even or odd parity with respect to the the $xy$-plane. In the Cartesian coordinate system, when $n+m$ is odd,
\begin{eqnarray}
\mathbf{P}^{(+x)}_{TE}=\mathbf{P}^{(-x)}_{TE}, \mathbf{P}^{(+y)}_{TE}=\mathbf{P}^{(-y)}_{TE}, \mathbf{P}^{(+z)}_{TE}=-\mathbf{P}^{(-z)}_{TE}\\* \nonumber
\mathbf{P}^{(+x)}_{TM}=-\mathbf{P}^{(-x)}_{TM}, \mathbf{P}^{(+y)}_{TM}=-\mathbf{P}^{(-y)}_{TM}, \mathbf{P}^{(+z)}_{TM}=\mathbf{P}^{(-z)}_{TM},
\label{eqn:mirsym_P_odd}
\end{eqnarray}
and when $n+m$ is even,
\begin{eqnarray}
\mathbf{P}^{(+x)}_{TE}=-\mathbf{P}^{(-x)}_{TE}, \mathbf{P}^{(+y)}_{TE}=-\mathbf{P}^{(-y)}_{TE}, \mathbf{P}^{(+z)}_{TE}=\mathbf{P}^{(-z)}_{TE}\\* \nonumber
\mathbf{P}^{(+x)}_{TM}=\mathbf{P}^{(-x)}_{TM}, \mathbf{P}^{(+y)}_{TM}=\mathbf{P}^{(-y)}_{TM}, \mathbf{P}^{(+z)}_{TM}=-\mathbf{P}^{(-z)}_{TM}.
\label{eqn:mirsym_P_even}
\end{eqnarray}
This allows a reduction in size of the A-matrix by a further factor of 4.
\section{The incident beam}
The optical tweezers that are used to drive micromachines are usually tightly focused laser beams. The beams my be linearly or circularly polarized (has spin angular momentum), and may or may not carry orbital angular momentum. For example, we use an $LG_{02}$ beam, which carries orbital angular momentum of $2\hbar$ per photon, to trap and rotate the microrotors such as the one in figure \ref{fig:cross}a. 

We calculate the incident field at each dipole using the vector spherical wave function expansion (VSWF) 
\begin{equation}
\mathbf{E}_\mathrm{inc} = \sum^{\infty}_{n=1}\sum^{n}_{m=-n} a_{nm} \mathbf{M}^{(3)}_{nm}(kr) + b_{nm} \mathbf{N}^{(3)}_{nm}(kr),
\label{eqn:E_inc}
\end{equation}
where $k$ is the wave vector, $r$ is the dipole position in spherical coordinates, \textit{n} is the radial mode index, \textit{m} is the azimuthal mode index and $\mathbf{M}^{(3)}_{nm}$ \& $\mathbf{N}^{(3)}_{nm}$ are regular VSWFs~\cite{Mishchenko2000,Nieminen2003a}; $a_{nm}$ and $b_{nm}$ are incident coefficients for the illuminating beam calculated using incident beam functions from \cite{Nieminen2007a} which uses the point matching method. If it sufficient to terminate the multipole expansion at $n = N_{max}$ where $N_{max} = ka + 3\sqrt[3]{ka}$~\cite{Brock2001}.
\section{Formulating the \emph{T}-matrix with point matching}
\label{Tmatrix}
Here, we present the point matching method as an alternative to an existing T-matrix method with DDA~\cite{Mackowski2002}. The point matching method that we implemented involved matching the fields due to the dipoles with the fields calculated from vector spherical wave function expansion and at multiple points. The number of points should be such that the linear system of equations are exactly or over determined i.e. the number of equations are the same or greater than the number of unknowns. 

Once we have the dipole moments, the $\mathbf{E}$-field at any point $\mathbf{r}$ (relative to the origin) can be calculated by adding up the contributions from each dipole using the electric dipole field equation from section 9.2 of \cite{Jackson1998},
\begin{equation}
\label{eqn:Jackson_dipole_E}
\mathbf{E}(k\mathbf{r}) = \frac{1}{4\pi\epsilon_0} \left\lbrace k^2(\mathbf{\hat{r}} \times \mathbf{p}) \times \mathbf{\hat{r}} \frac{e^{\mathrm{i}kr}}{r} + [3\mathbf{\hat{r}}(\mathbf{\hat{r}}\cdot\mathbf{p})-\mathbf{p}] \left(\frac{1}{r^3}-\frac{\mathrm{i}k}{r^2}\right) e^{\mathrm{i}kr}\right\rbrace, 
\end{equation}
In the far field, the $1/r^2$ and $1/r^3$ terms can be ignored as their contributions dimish rapidly with distance:
\begin{equation}
\label{eqn:Jackson_dipole_E_FF}
\mathbf{E}(k\mathbf{r}) = \frac{1}{4\pi\epsilon_0} \left\lbrace k^2(\mathbf{\hat{r}} \times \mathbf{p}) \times \mathbf{\hat{r}} \frac{e^{\mathrm{i}kr}}{r} \right\rbrace. 
\end{equation}
\subsection{Near field point matching}
\label{sect:nearfield}
\begin{figure}[ht!]
\centerline{%
(a)\includegraphics[height=6cm]{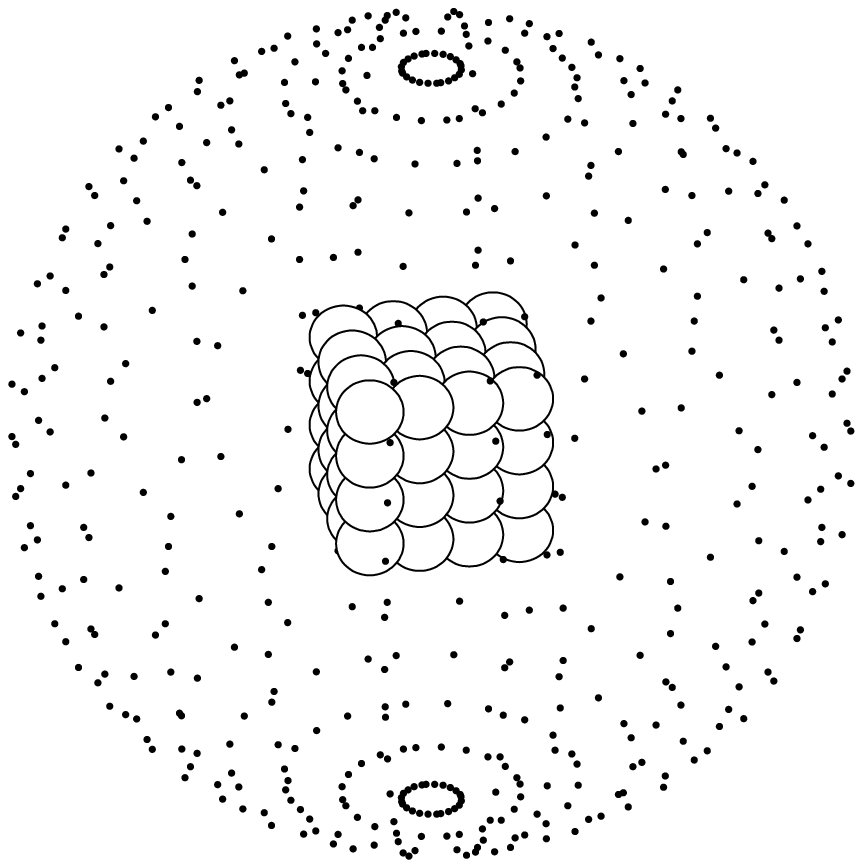}
\hfill
(b)\includegraphics[height=4cm]{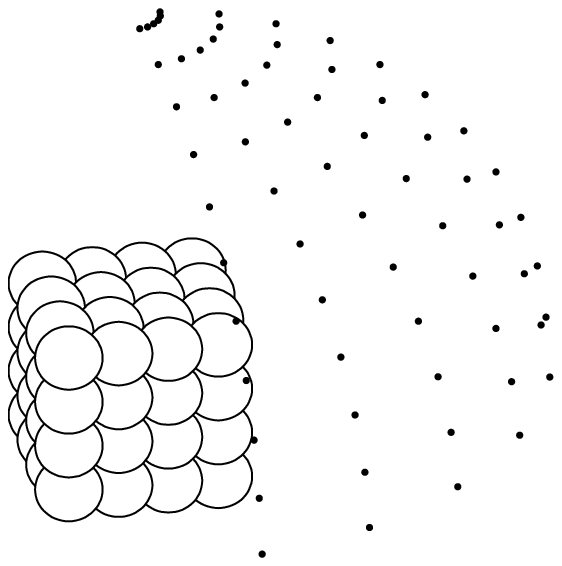}}
\caption{(a)~Near field matching and (b)~octant near field matching for a dipole model of a cube.}
\label{fig:NF_PM}
\end{figure}

The near field calculated via DDA (\ref{eqn:Jackson_dipole_E}) can be matched to scattered field (\ref{eqn:E_sca}) calculated via the VSWFs at points around the scatterer (figure \ref{fig:NF_PM}). Since we have the dipole moments, $P_j$, we can derive a `field matrix', $\mathbf{F}_{ij}$, derived from (\ref{eqn:Jackson_dipole_E}) such that
\begin{equation}
\label{eqn:DDA_nearfield}
\mathbf{E}^{(DDA)}_{i}=\sum^{N_{PM}}_{i=1}\mathbf{F}_{ij}\mathbf{P}_j,
\end{equation}
where $N_{PM}$ is the number of point to be matched, $j$ is the index for the dipole, $i$ is the index for the near field point and each element $F_{ij}$ is a $3\times 3$ tensor. Using the vector identity $(\mathbf{a}\times\mathbf{b})\times\mathbf{c} = \mathbf{b}(\mathbf{a}\cdot\mathbf{c})-\mathbf{c}(\mathbf{a}\cdot\mathbf{b})$, the cross product terms in (\ref{eqn:Jackson_dipole_E}) become
\begin{eqnarray}
\label{eqn:dipole_pr}
(\mathbf{\hat{r}}\times\mathbf{p})\times\mathbf{\hat{r}} &=& \mathbf{p}(\mathbf{\hat{r}}\cdot\mathbf{\hat{r}})-\mathbf{\hat{r}}(\mathbf{\hat{r}}\cdot\mathbf{p})\\ \nonumber
&=&\mathbf{p}-\mathbf{\hat{r}}(\mathbf{\hat{r}}\cdot\mathbf{p}).
\end{eqnarray}
Substituting the result of (\ref{eqn:dipole_pr}) into (\ref{eqn:Jackson_dipole_E}) we obtain 
\begin{equation}
\label{eqn:AEp}
\mathbf{E} = \frac{e^{\mathrm{i}kr}}{r}\left[k^2(\mathbf{\hat{r}}\mathbf{\hat{r}}-\mathbf{1}_3) + \frac{(\mathrm{i}kr-1)}{r^2}(3\mathbf{\hat{r}}\mathbf{\hat{r}}-\mathbf{1}_3)\right]\mathbf{p},
\end{equation}
which is can also be used to formulate the A-matrix (\ref{eqn:A_offdiag}). Thus the field matrix in (\ref{eqn:DDA_nearfield}) is
\begin{equation}
\label{eqn:NF_matrix}
\mathbf{F}_{ij} = \frac{\exp(\mathrm{i}kr_{jk})}{r_{jk}} \left[ k^2(\hat{r}_{jk}\hat{r}_{jk}-\mathbf{1}_3)+ \frac{\mathrm{i}kr_{jk}-1}{r_{jk}} (3\hat{r}_{jk}\hat{r}_{jk}-\mathbf{1}_3) \right],
\end{equation}
where $\hat{r}_{jk}$ is the unit vector between a dipole and a matched point. ${F}_{ij}$ can be precalculated and used when required.

Rotational (\ref{eqn:rotsym_P_cart}) and mirror (\ref{eqn:mirsym_P_odd}) symmetry optimizations may also be applied to (\ref{eqn:DDA_nearfield}) such that we only need to calculate the fields of the matched points in one octant:
\begin{equation}
\label{eqn:DDA_nearfield_oct}
\mathbf{E}^{(DDA,oct)}_i=\sum^{N_{PM}/8}_{i=1}\mathbf{F}^{(oct)}_{ij}\mathbf{P}^{(oct)}_j,
\end{equation}
where only an the fields for an octant are calculated (although all the dipole moments are used). Thus reduces the number of equations by a factor of $64$ compare to (\ref{eqn:DDA_nearfield}).
Now, since the VSWF expansion of the scattered field is
\begin{equation}
\mathbf{E}_\mathrm{sca} = \sum^{\infty}_{n=1}\sum^{n}_{m=-n} p_{nm} \mathbf{M}^{(1)}_{nm}(kr) + q_{nm} \mathbf{N}^{(1)}_{nm}(kr).
\label{eqn:E_sca}
\end{equation} 
we can solve for the scattering coefficients
\begin{eqnarray}
 &p_{nm} = \mathbf{M}_{nm}^{(1)}(kr)/\mathbf{E}^{(DDA)}_{TE,nm},\\ \nonumber
 &q_{nm} = \mathbf{N}_{nm}^{(1)}(kr)/\mathbf{E}^{(DDA)}_{TM,nm}.
\end{eqnarray}
As we cycle through each combination of $n$ and $m$, we obtain the solutions for the scattering coefficients $p_{nm}$ and $q_{nm}$ which represent coupling between the $n$ and $m$ incident and scattered modes; $p_{nm}$ and $q_{nm}$ together make up one column of the \emph{T}-matrix at a time. 
\subsection{\label{farfield}Far field matching}
In this method we match the VSWF expansion of the scattered field (\ref{eqn:E_sca}) with that calculated via DDA in far field. We calculate the DDA field in a way similar to (\ref{eqn:DDA_nearfield}) but instead, calculating the far field
\begin{equation}
\label{eqn:DDA_farfield}
kr\mathbf{E}^{(DDA)}_{i}=\sum^{N_{PM}}_{i=1}\mathbf{F}_{ij}\mathbf{P}_j,
\end{equation}
where field matrix, $\mathbf{F}_{ij}$, from (\ref{eqn:NF_matrix}) is multiplied by $kr$ and with the latter term omitted to give
\begin{equation}
\label{eqn:FF_matrix}
\mathbf{F}_{ij} = \exp(\mathrm{i}kr_{jk}) k^3(\hat{r}_{jk}\hat{r}_{jk}-\mathbf{1}_3).
\end{equation}
Now the VSWFs in the expansion of the scattered field (\ref{eqn:E_sca}) are
\begin{equation}
\label{eqn:Mnm}
\mathbf{M}^{(1)}_{nm}(kr) = \mathbf{N}_nh^{(1)}_n(kr)\mathbf{C}_{nm}(\theta,\phi), 
\end{equation}
\begin{eqnarray}
\label{eqn:Nnm}
\mathbf{N}^{(1)}_{nm}(kr) = &&\frac{h^{(1)}_n(kr)}{krN_n}\mathbf{P}_{nm}(\theta,\phi) +\nonumber\\
&&N_n\left(h^{(1)}_{n-1}(kr)-\frac{nh^{(1)}_n(kr)}{kr}\right)\mathbf{B}_{nm}(\theta,\phi)
\end{eqnarray}
where $N_n=1/\sqrt{n(n+1)}$, $h^{(1,2)}$ are spherical Hankel functions and $\mathbf{B}_{nm}$, $\mathbf{C}_{nm}$ \& $\mathbf{P}_{nm}$ are vector spherical harmonics: 
\begin{eqnarray}
\mathbf{B}_{nm}(\theta,\phi) &&=r\nabla\gamma^m_n(\theta,\phi)\nonumber\\
&&=\nabla\times \mathbf{C}^n_m(\theta,\phi)\nonumber\\
&&=\hat{\theta} \frac{\partial}{\partial\theta}\gamma^m_n(\theta,\phi)+\hat{\phi} \frac{im}{\sin\theta}\gamma^m_n(\theta,\phi),
\end{eqnarray}
\begin{eqnarray}
\mathbf{C}_{nm}(\theta,\phi) &&=\nabla\times(r\gamma^m_n\theta,\phi)\nonumber\\
&&=\hat{\theta} \frac{im}{\sin\theta}\gamma^m_n(\theta,\phi)-\hat{\phi} \frac{\partial}{\partial\theta}\gamma^m_n(\theta,\phi),
\end{eqnarray}
\begin{eqnarray}
\label{eqn:P}
\mathbf{P}_{nm}(\theta,\phi) =\hat{r}\gamma^m_n(\theta,\phi)
\end{eqnarray}
where $\gamma^m_n(\theta,\phi)$ is the normalised scalar spherical harmonics. $\mathbf{M}^{(1)}_{nm}$ and $\mathbf{N}^{(1)}_{nm}$ are the TE and TM multipole fields respectively. In the farfield limit~\cite{Mishchenko1991,Nieminen2003a} the VSWFs (\ref{eqn:Mnm}) \& (\ref{eqn:Nnm}) become 
\begin{eqnarray}
 &&\mathbf{M}_{nm}^{(1)}(kr) = \frac{N_n}{kr}(-\mathrm{i})^{n+1} \exp(\mathrm{i}kr) \mathbf{C}_{nm}(\theta,\phi)\\
 &&\mathbf{N}_{nm}^{(1)}(kr) = \frac{N_n}{kr}(-\mathrm{i})^{n+1} \exp(\mathrm{i}kr) \mathbf{B}_{nm}(\theta,\phi).
\end{eqnarray}
The scattered can be further simplified since $\mathop {\lim }\limits_{r \to \infty } \exp(\mathrm{i}kr)\rightarrow 1$,
\begin{eqnarray}
\label{eqn:MN_CB}
 &&\mathbf{M}_{nm}^{(1)}(kr) = \frac{N_n}{kr}(-\mathrm{i})^{n+1} \mathbf{C}_{nm}(\theta,\phi)\\
 &&\mathbf{N}_{nm}^{(1)}(kr) = \frac{N_n}{kr}(-\mathrm{i})^{n+1} \mathbf{B}_{nm}(\theta,\phi),
\end{eqnarray}
Now the far field can also be calculated using
\begin{equation}
\label{eqn:sum_BCP}
 kr \mathbf{E}(\theta,\phi)=\sum_{nm} b_{nm}\mathbf{B}_{nm} + c_{nm}\mathbf{C}_{nm} + a_{nm}\mathbf{P}_{nm},
\end{equation}
where the coefficients $b_{nm}$ \& $c_{nm}$ can be calculated (trapezoidal integration) using 
\begin{equation}
b_{nm} = \frac{\int kr\mathbf{E}\cdot\mathbf{B}_{nm} d\sigma}{\int \mathbf{B}_{nm}\cdot\mathbf{B}_{nm} d\sigma},
\end{equation}
\begin{equation}
c_{nm} = \frac{\int kr\mathbf{E}\cdot\mathbf{C}_{nm} d\sigma}{\int \mathbf{C}_{nm}\cdot\mathbf{C}_{nm} d\sigma},
\end{equation}
where $kr\mathbf{E}$ is calculated via DDA; this is where the point matching occurs. The last term in (\ref{eqn:sum_BCP}) can be ignored because the radial component (\ref{eqn:P}) is negligible in the far field. If we then match the terms in (\ref{eqn:E_sca}), (\ref{eqn:sum_BCP}) \& (\ref{eqn:MN_CB}) the scattering coefficients can be determined:
\begin{equation}
 p_{nm} = \frac{b_{nm}}{N_n (-\mathrm{i})^{n+1}},
\end{equation}
\begin{equation}
 q_{nm} = \frac{a_{nm}}{N_n (\mathrm{i})^n}.
\end{equation}
So, as with the near field matching, we cycle through the $n$ and $m$ modes and insert $p_{nm}$ vertically concatenated with $q_{nm}$ into the appropriate column of the \emph{T}-matrix for the given mode.
\section{Mode redundancy and the order of discrete rotational symmetry}
We can also exploit the azimuthal dependence of $\exp(\mathrm{i}m\phi)$ of each individual VSWF. The discrete rotational symmetry effectively provides a periodic boundary condition, determining the periodicity with respect to the azimuthal angle $\phi$. According Floquet's theorem\cite{Russell1983}, in the case of the plane wave diffracting from a periodic grating, the wave vectors of the scattered modes are
\begin{equation}
k_{scat} = k_{inc} + ip, \; i = ...-2,-1,0,1,2,...
\end{equation}
where $p$ is the order of discrete rotational symmetry of the scatterer. The spherical wave equivalent is such that the azimuthal modes of the scattered light that couple to a given incident mode is
\begin{equation}
m_{scat} = m_{inc} + ip, \; i = ...-2,-1,0,1,2,....
\end{equation}
For a structure with four-fold discrete rotational symmetry, $p = 4$, thus $m_{sca} = m_{inc}, m_{inc}\pm 4, m_{inc}\pm 8, ...$. The number of unknowns in the linear system is reduced by a factor of 4. The calculation time saving for this scheme is about 2 to 3 orders of magnitude.
\section{Mode-reduced T-matrix}
This scheme may be deemed in breach of the true purpose of the \emph{T}-matrix i.e. it should be good for any illumination at the wavelength for which it was formulated. However, if we know that the only beam that we are going to use is, say the $LG_{02}$ and that the scatterer e.g. figure \ref{fig:cross}a only spins along the beam axis, and stays on the beam axis, then the only relevant scattering modes when calculating the \emph{T}-matrix would be $m=1,3$. 

Bearing in mind that we 'move' the scatterer along the beam axis to find the equilibrium position by calculating the axial force~\cite{Nieminen2007a}; we perform repeated calculations with the mode-reduced \emph{T}-matrix without sacrificing any accuracy. Formulating the mode-reduced \emph{T}-matrix, in most cases, takes a fraction of the time taken for the full-mode \emph{T}-matrix.
\section{Results}
\begin{figure}[h!t]
\centering
a)\includegraphics[height=6cm]{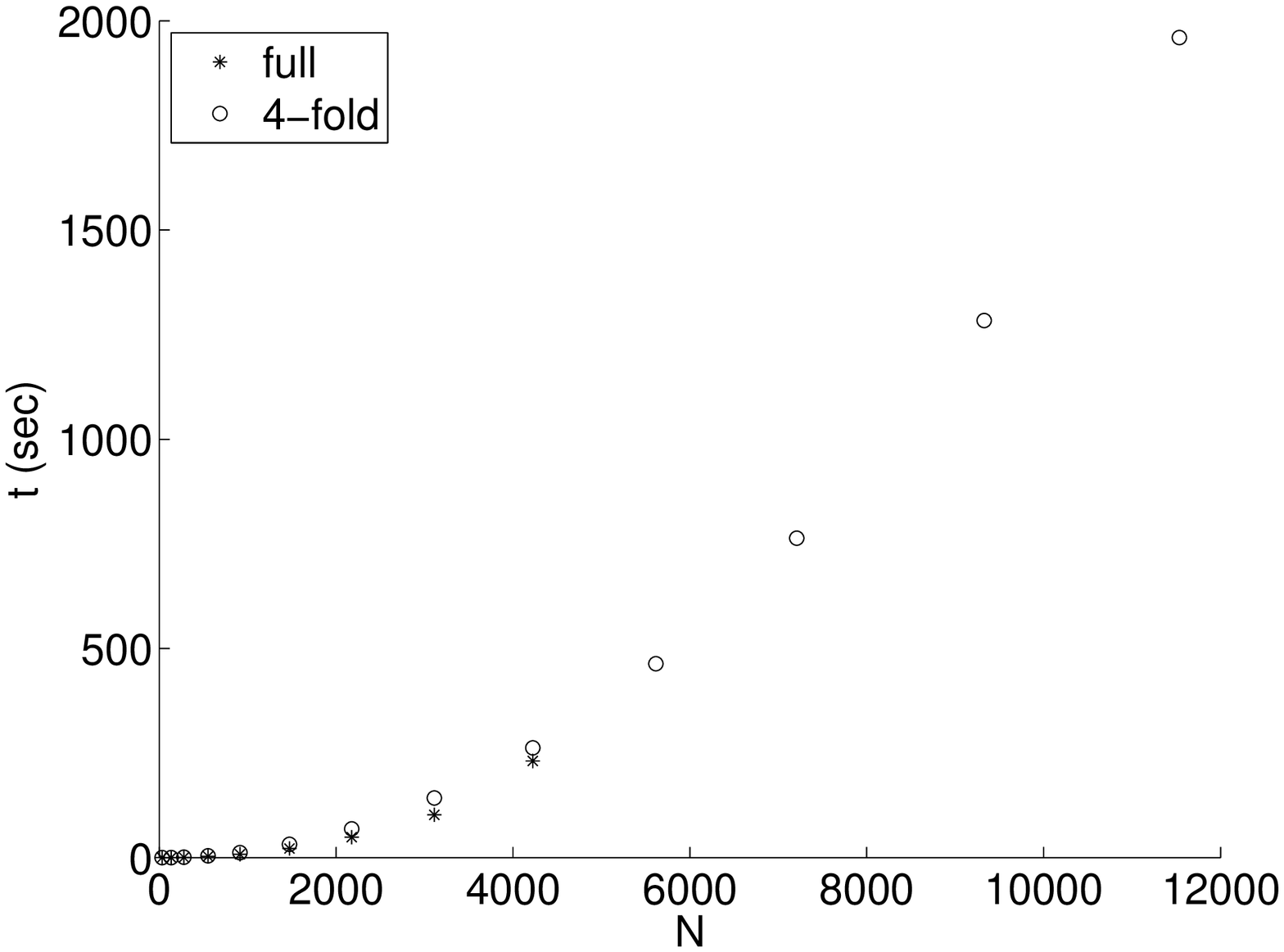}\\
b)\includegraphics[height=6cm]{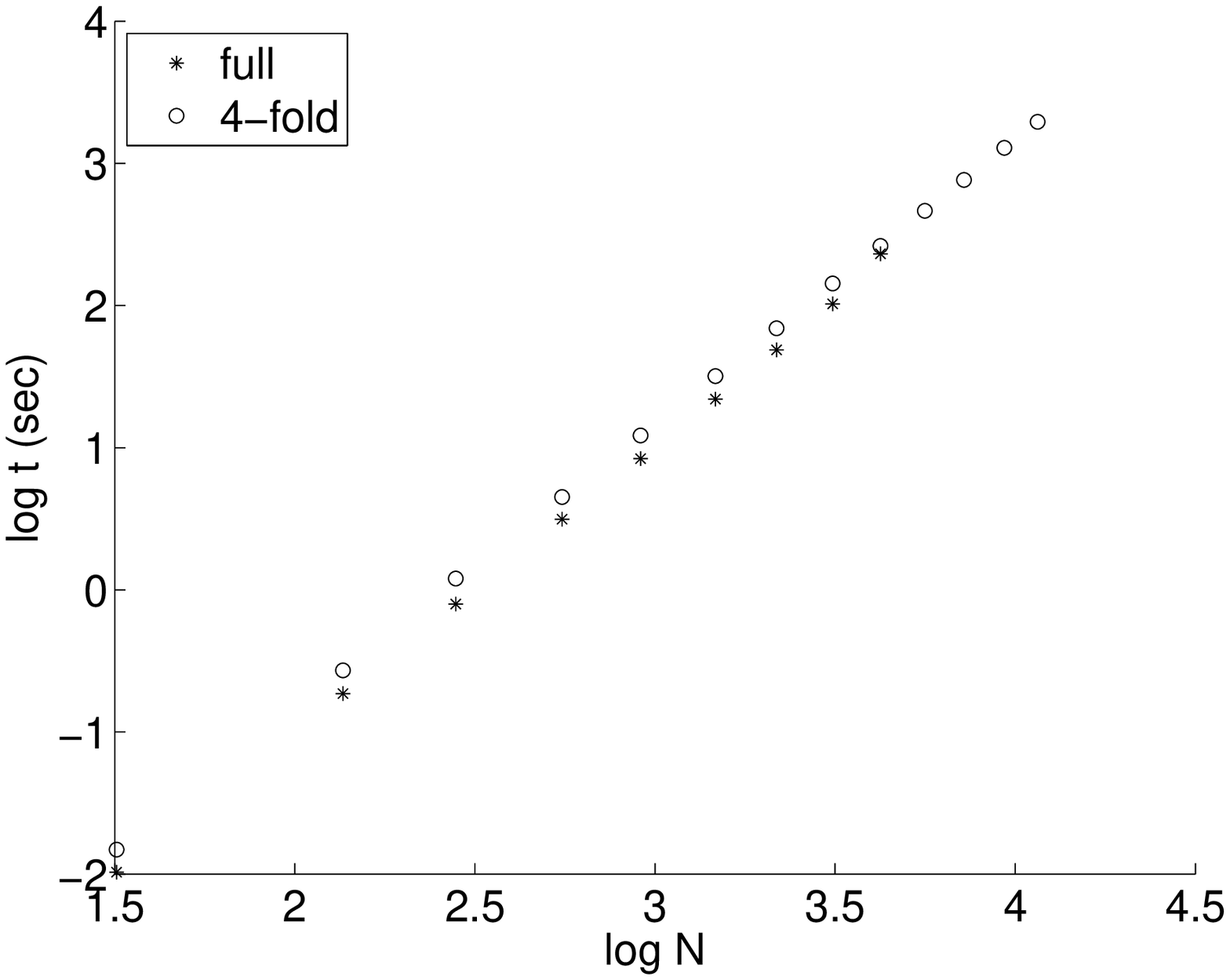}
\caption{a)Linear and b)log-log plots of the time required to assemble the interaction matrices for versus the number of dipoles (or equivalent for the 4-fold discrete rotational symmetric methodology}
\label{fig:calc_A}
\end{figure}
\begin{figure}[h!t]
\centering
a)\includegraphics[height=6cm]{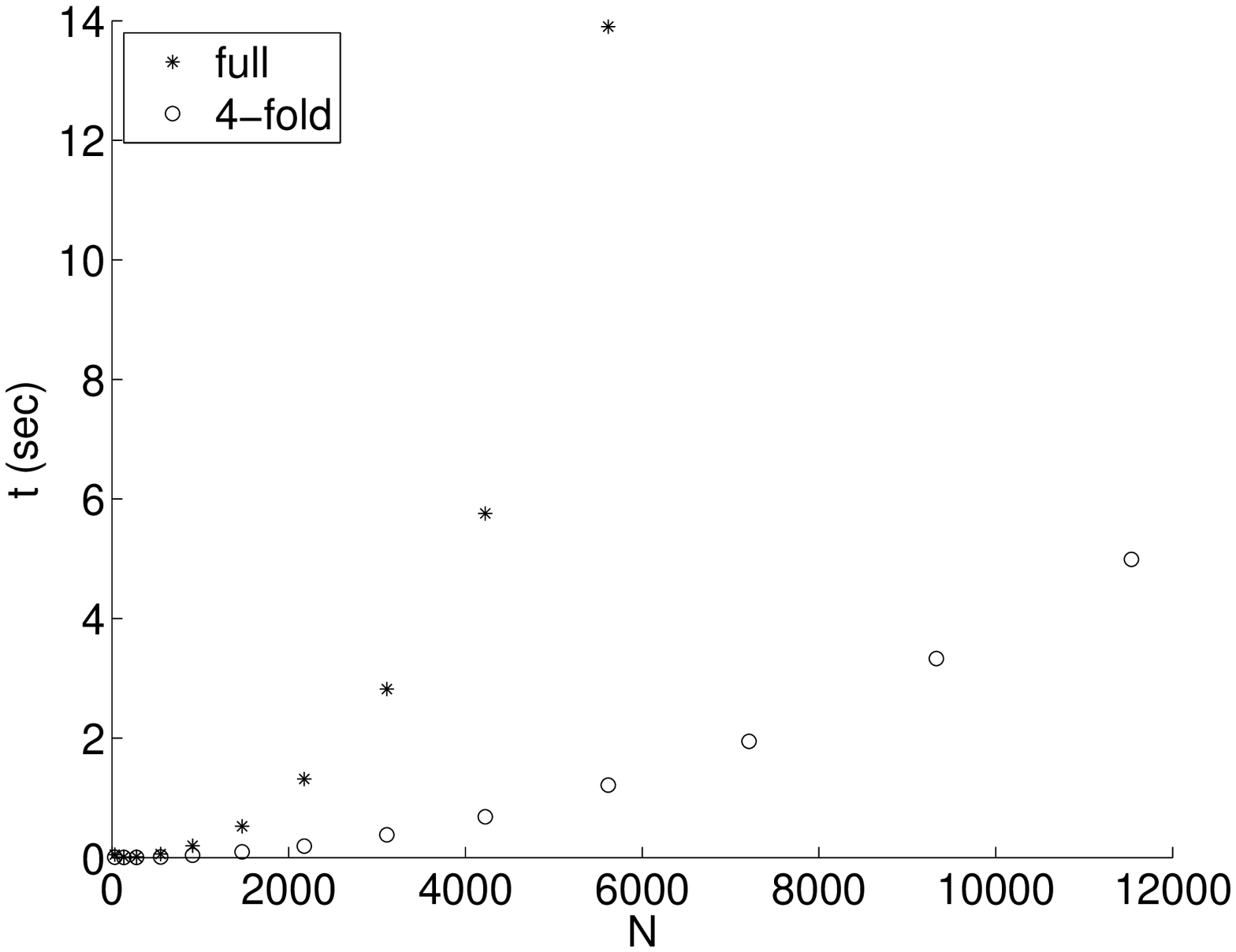}\\
b)\includegraphics[height=6cm]{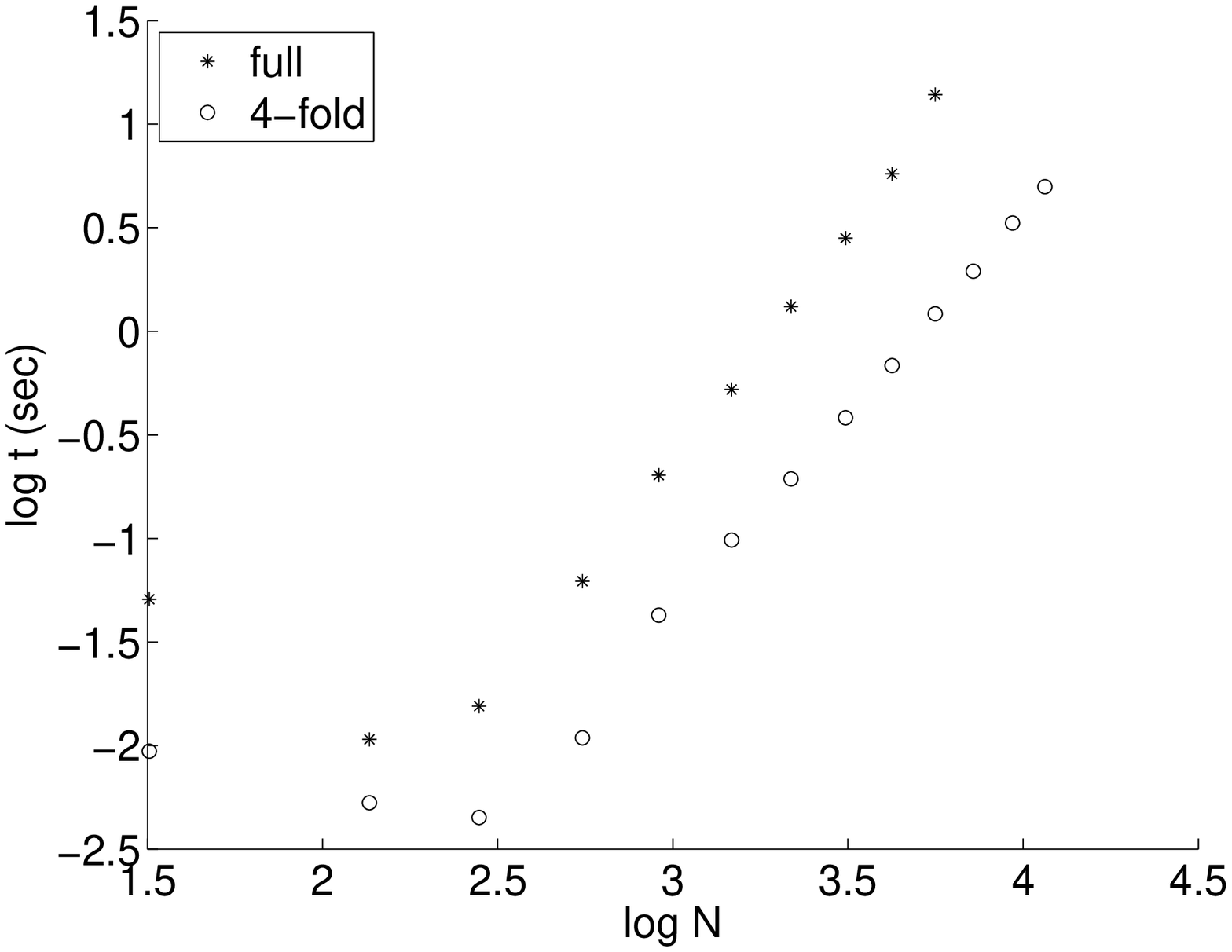}
\caption{a)Linear and b)log-log plots of the time required to solve for the dipole moments versus the number of dipoles}
\label{fig:calc_P}
\end{figure}
\begin{figure}[htp]
\centering
a)\includegraphics[height=4.1cm]{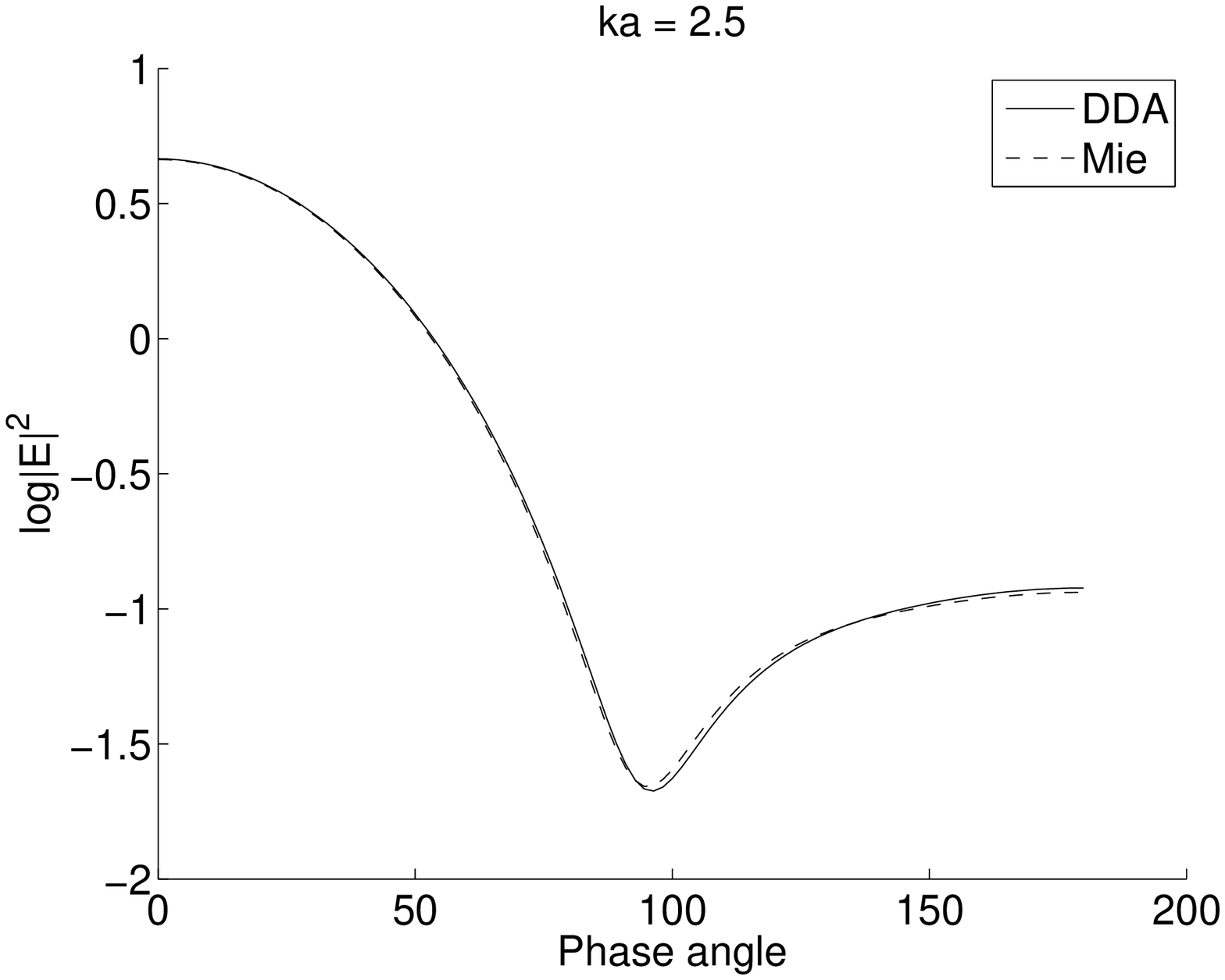}b)\includegraphics[height=4.1cm]{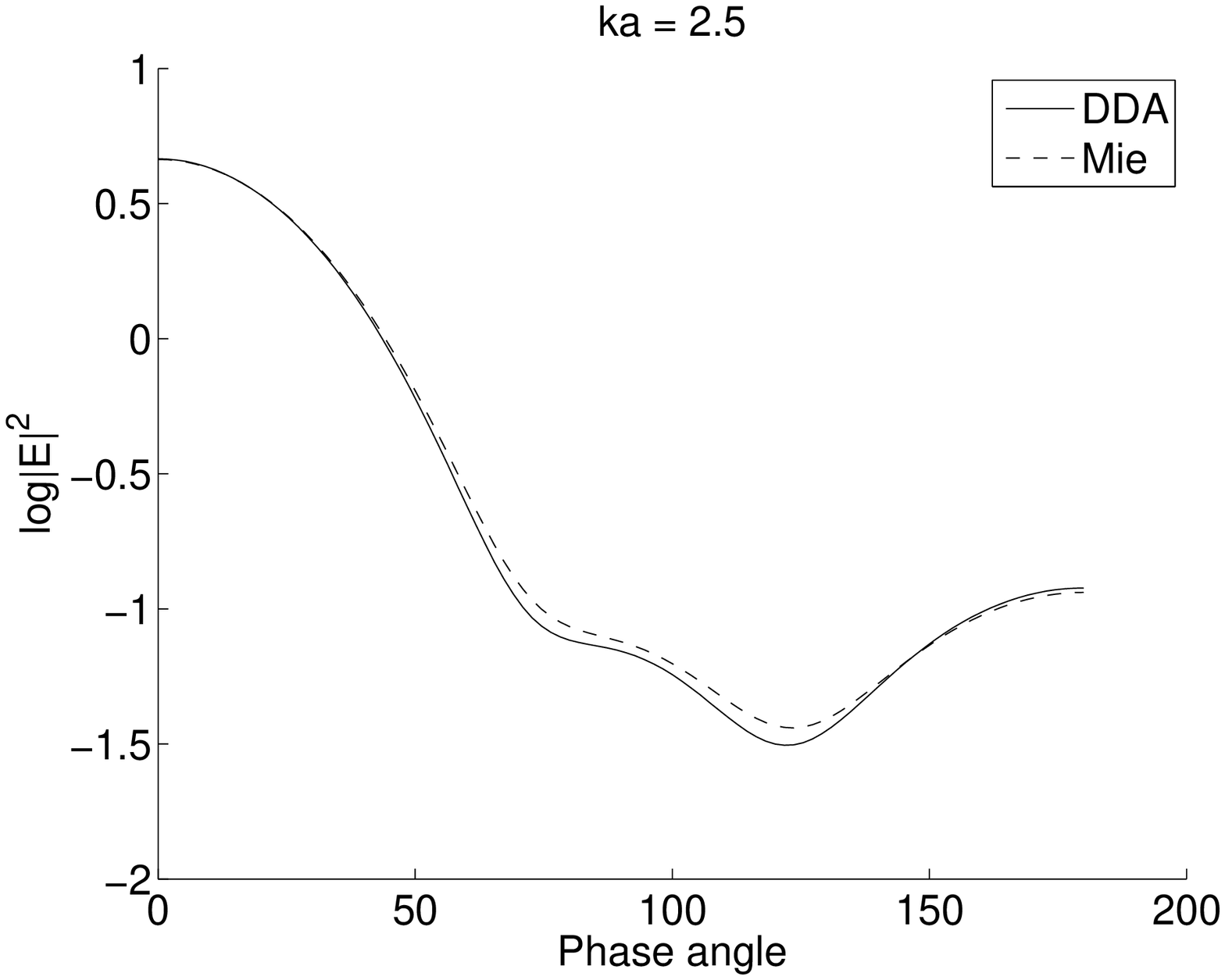}\\
c)\includegraphics[height=4.1cm]{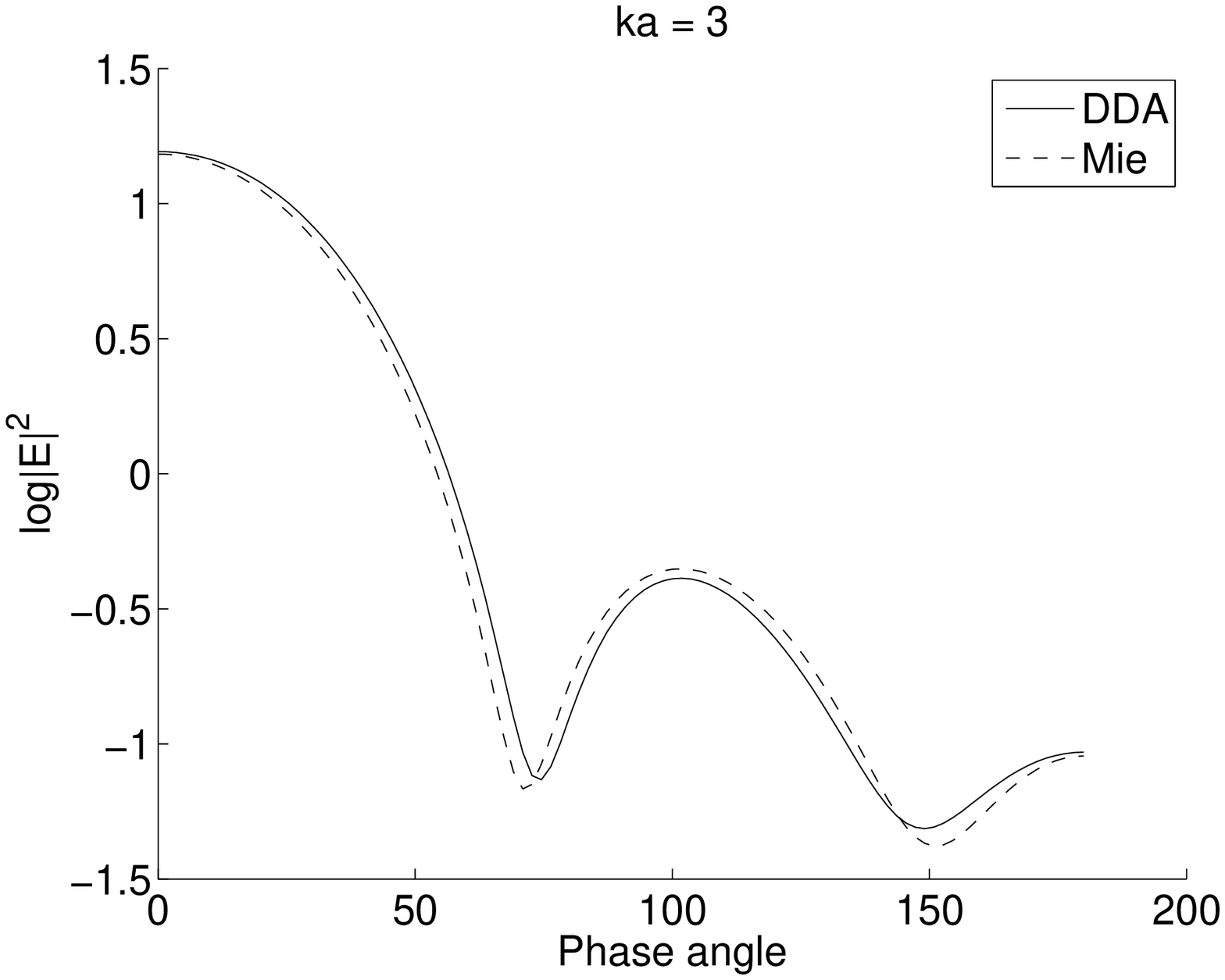}d)\includegraphics[height=4.1cm]{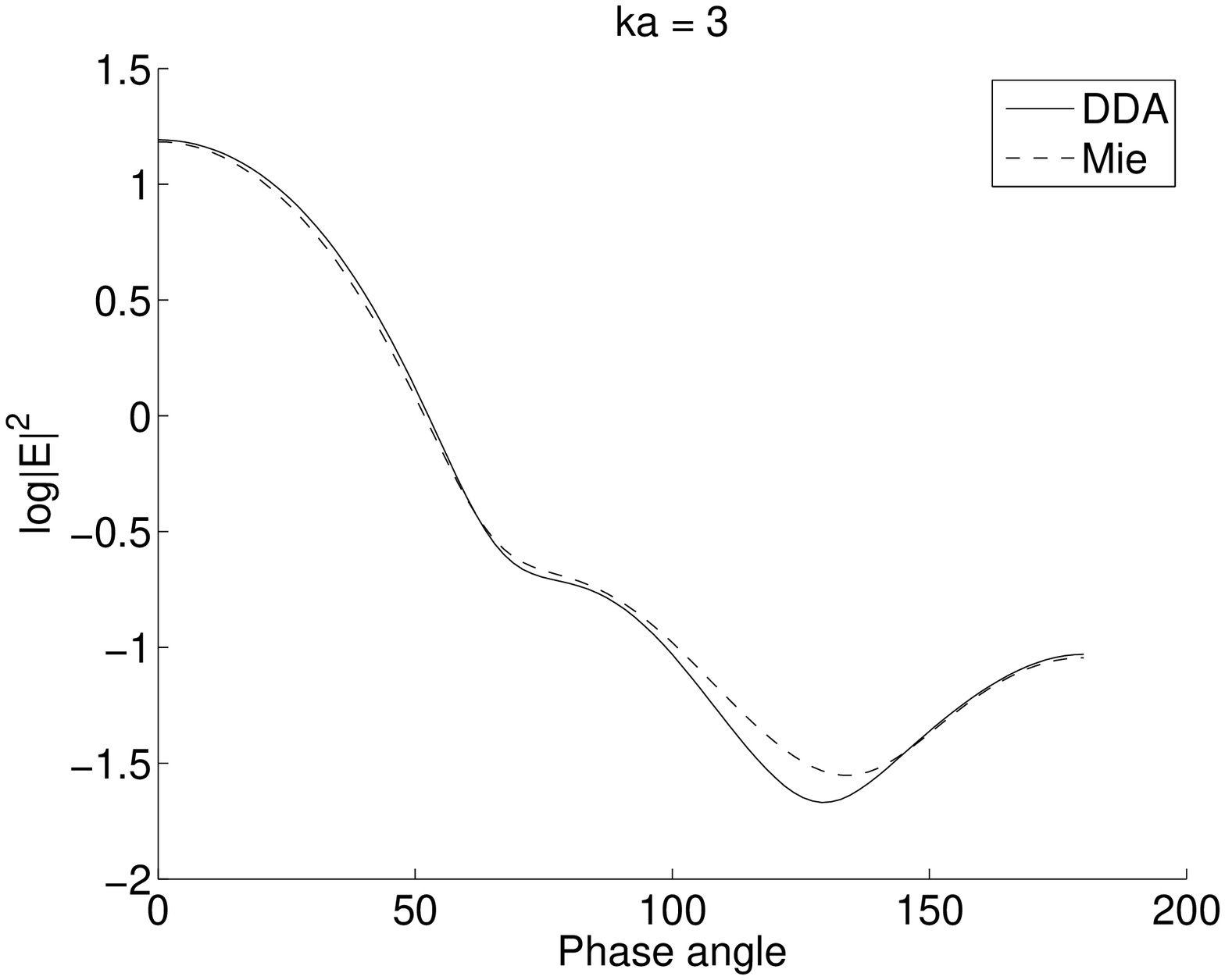}\\
e)\includegraphics[height=4.1cm]{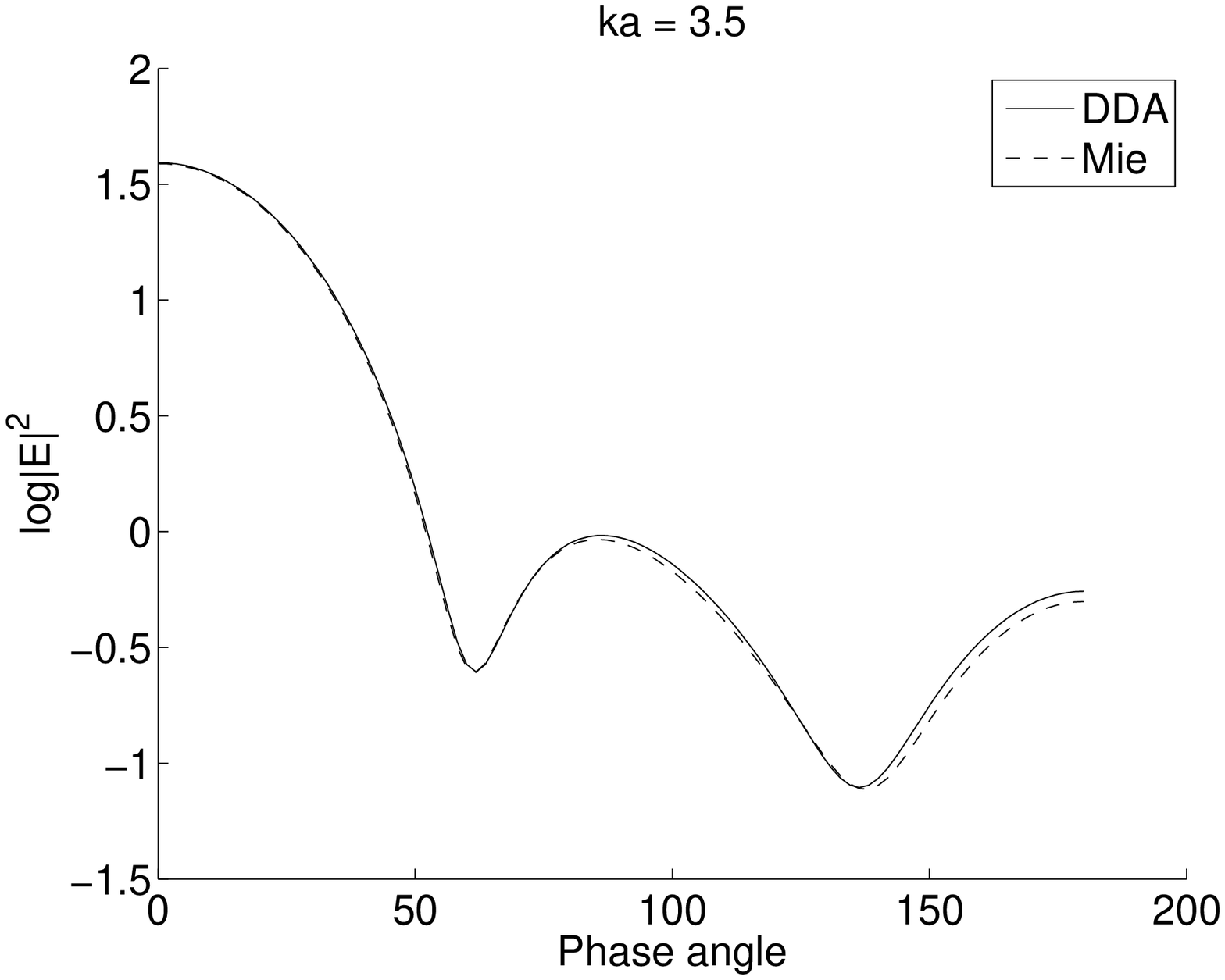}f)\includegraphics[height=4.1cm]{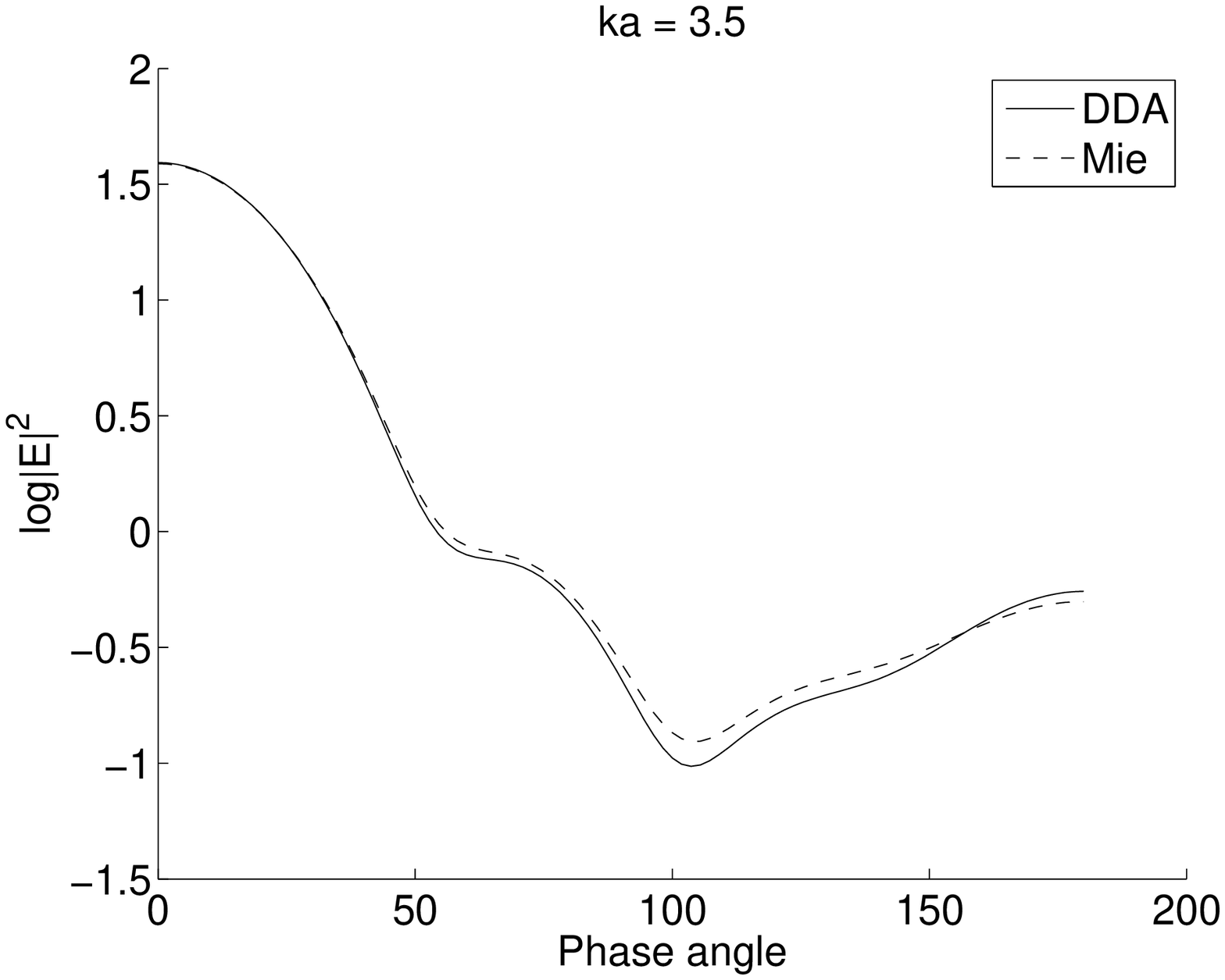}\\
g)\includegraphics[height=4.1cm]{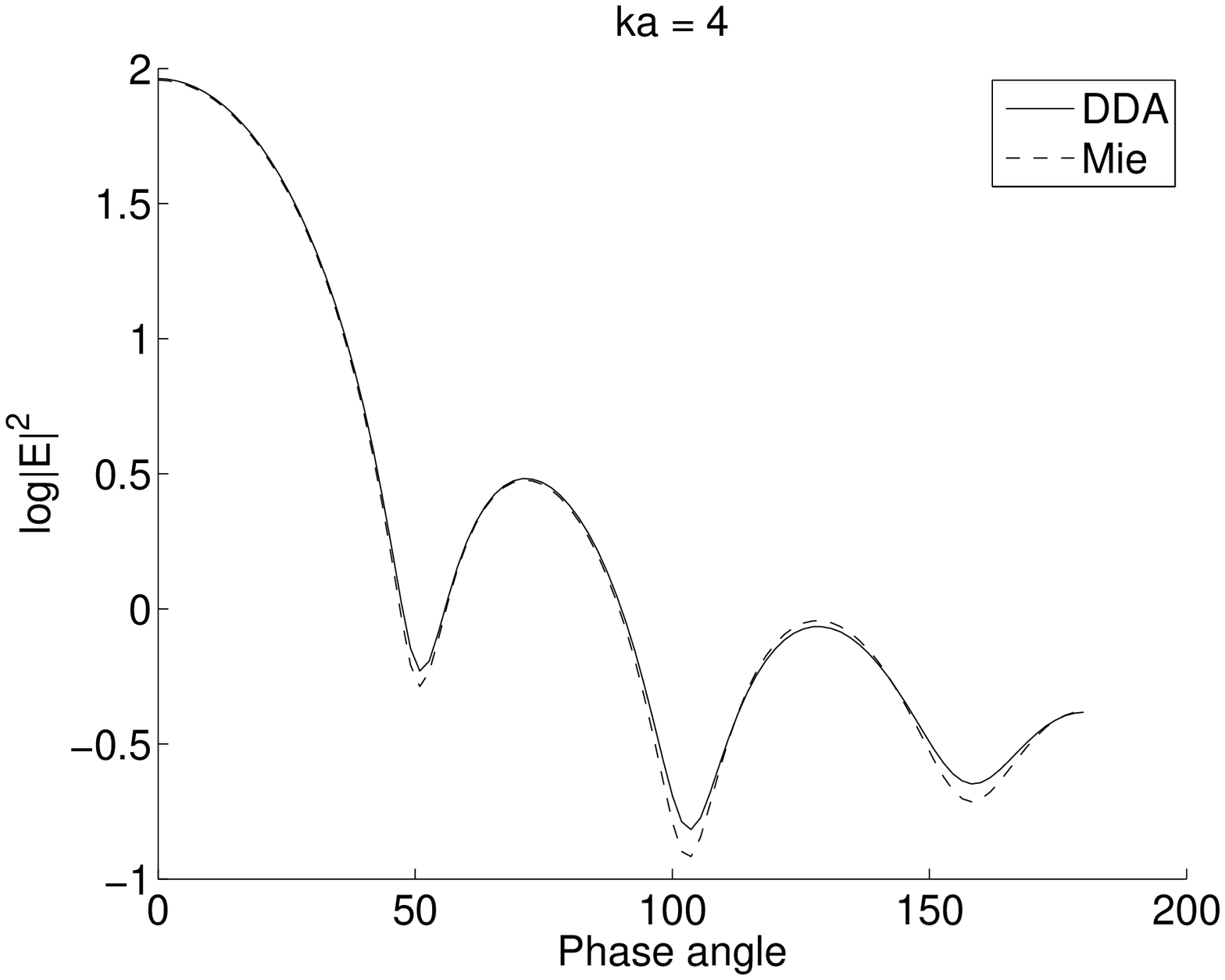}h)\includegraphics[height=4.1cm]{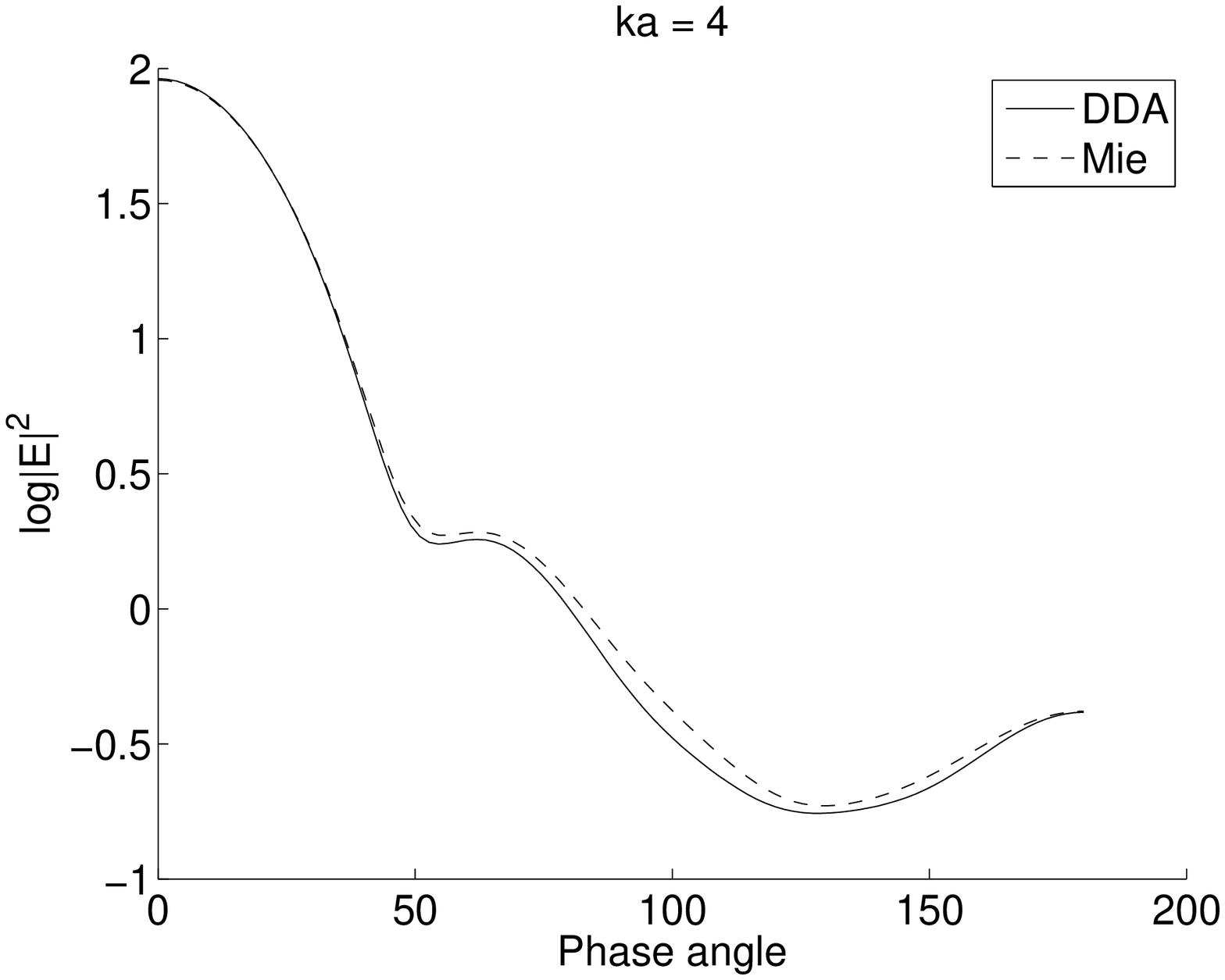}
\caption{The phase function plots on the left column are for the plane perpendicular to the ($\mathbf{E}$-field) polarization of the incident plane wave whereas those on the right column are for the parallel plane. The plots for varying size parameters (2.5, 3, 3.5 $\&$ 4) are from top to bottom. The relative refractive index is 1.33.}
\label{fig:NF_sphere_phasefn}
\end{figure}
\begin{figure}[htp]
\label{fig:WC_cube}
\centering
a)\includegraphics[height=5cm]{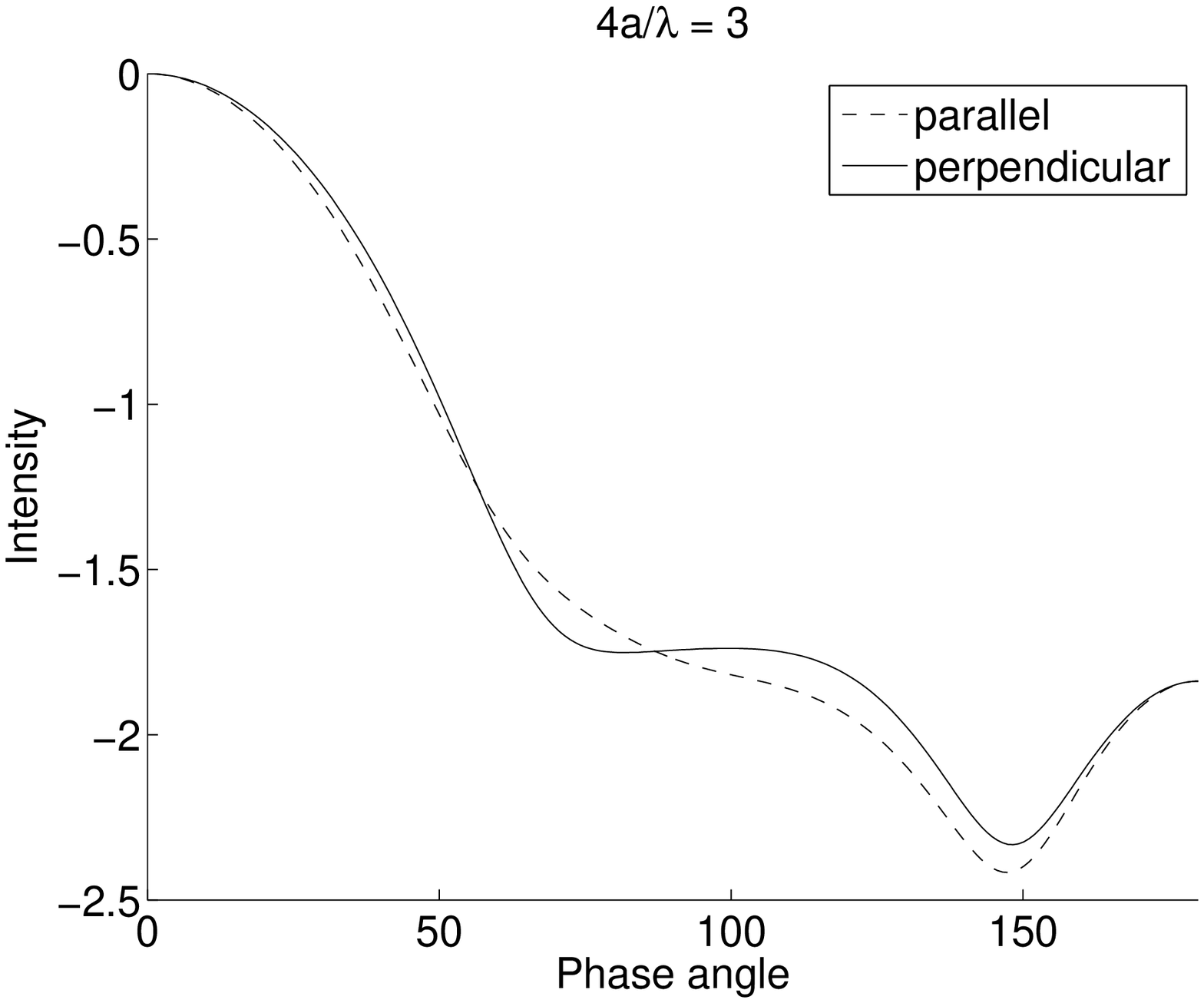} b)\includegraphics[height=5cm]{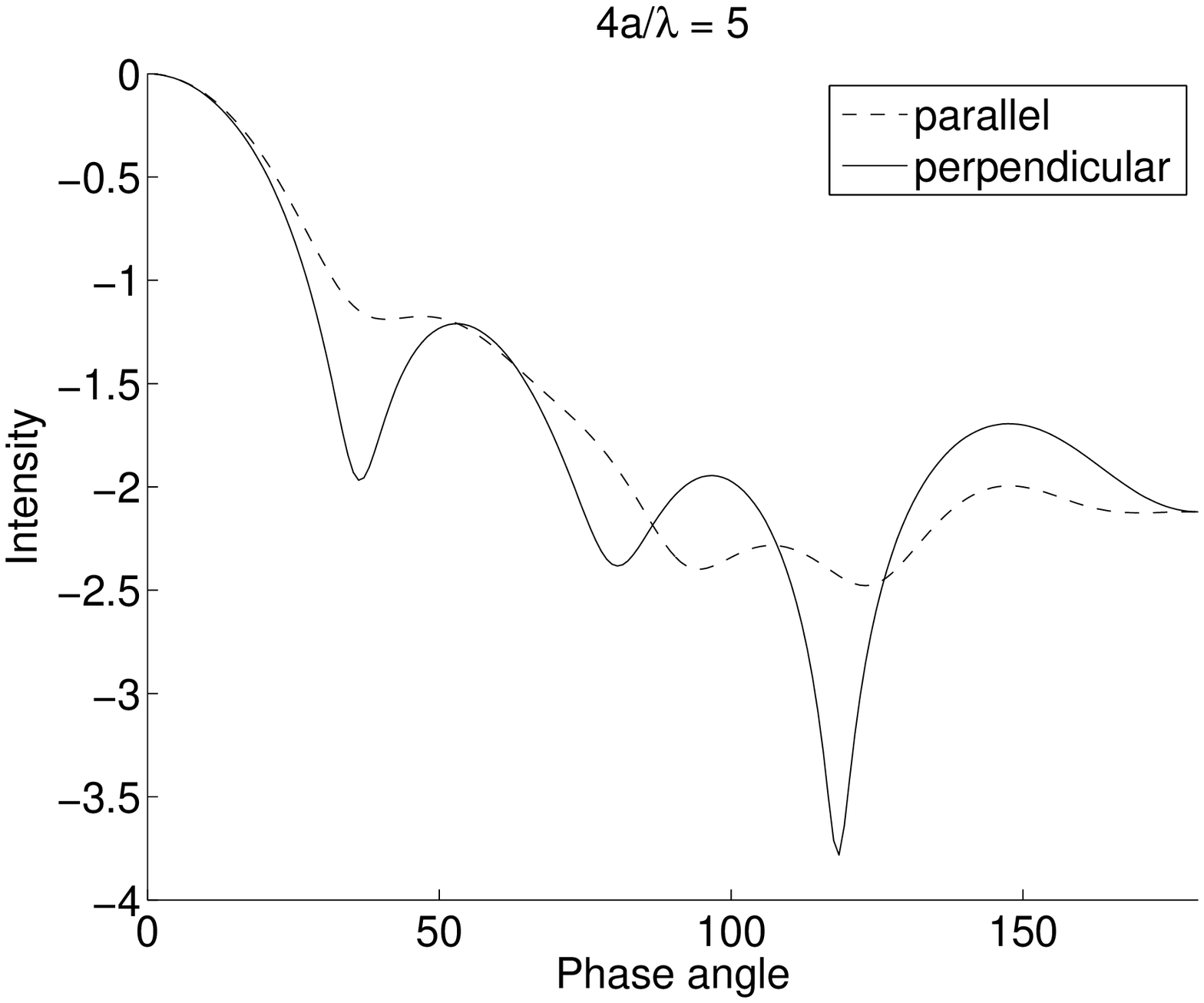}
\caption{The phase function for cubes with the refractive index of $1.5$ and widths of a)~$0.75 \lambda$ b)~$1.25 \lambda$.}
\end{figure}
\begin{figure}[htp]
\label{fig:force_torque}
\centering
a)\includegraphics[height=5cm]{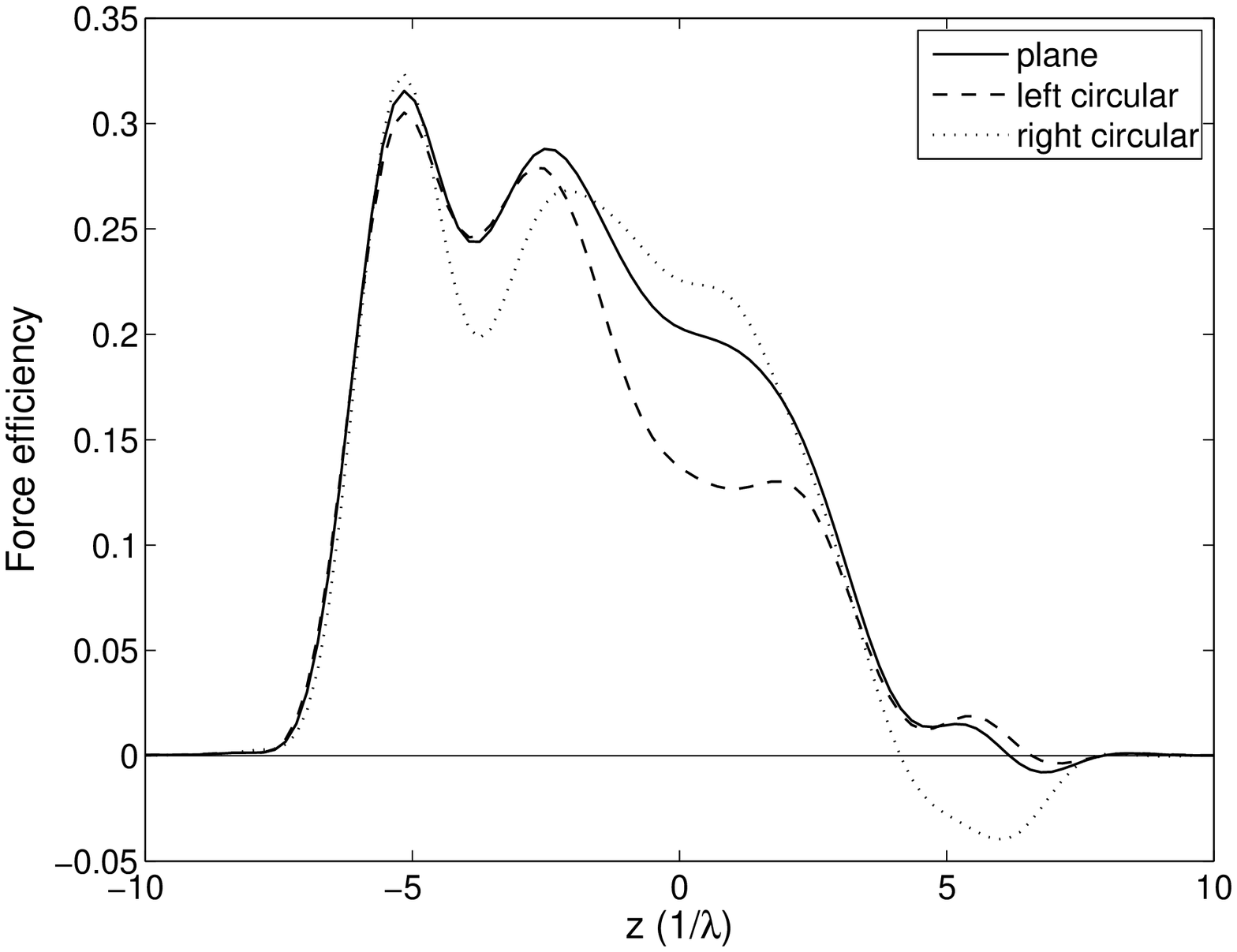}b)\includegraphics[height=5cm]{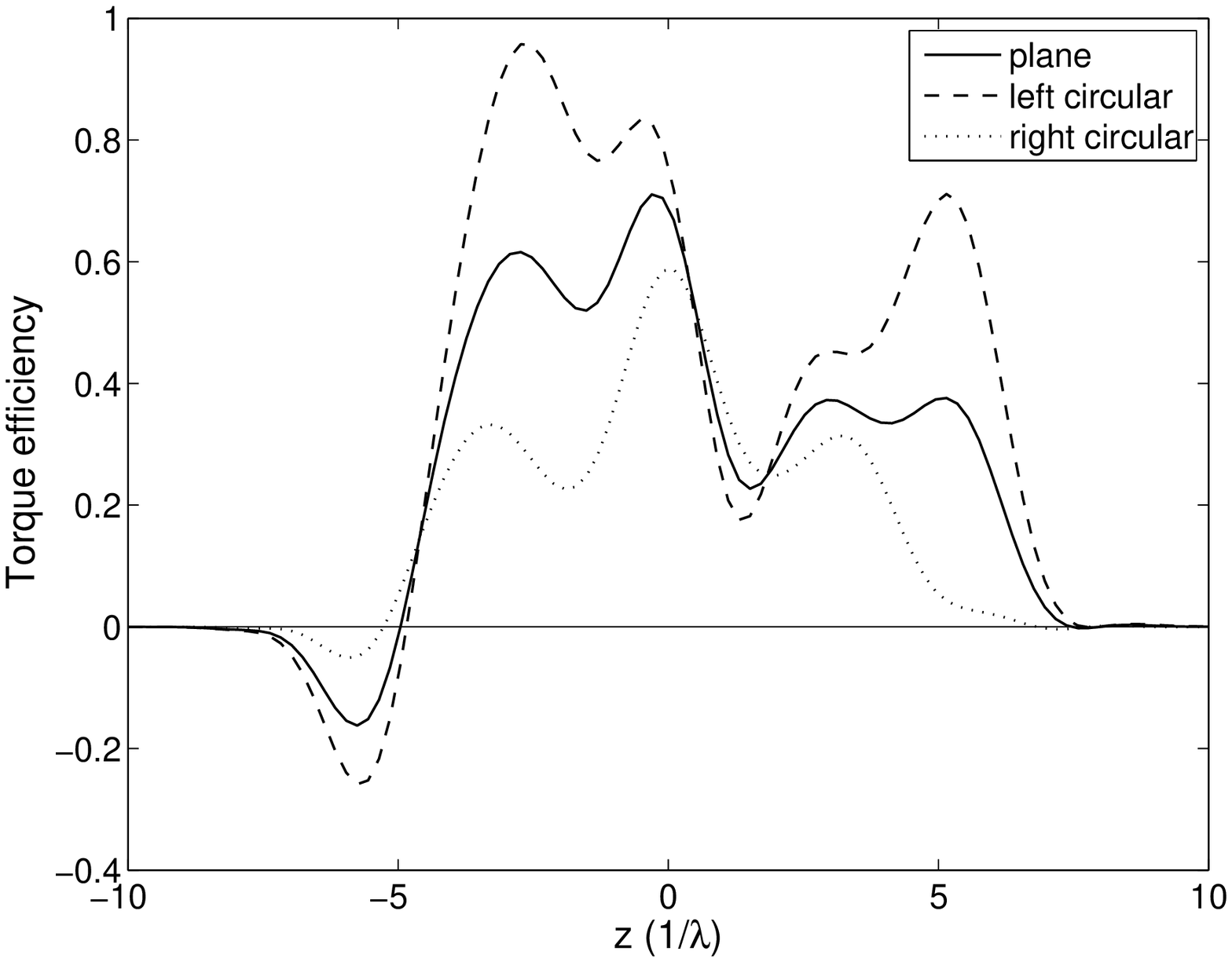}
\caption{The axial a)~force and b)~torque as a function of the displacement along the beam axis for an $LG_{02}$ beam with a $71.55^o$ convergence angle, for plane and circular polarisations.}
\end{figure}
\subsection{Symmetry optimization}
The rotational symmetry optimized DDA scheme produces results that are no different (well below round-off errors) to that of the standard DDA implementation; there is no appreciable difference that would be visible when plotting comparison data for say the phase function or the extinction. Instead, we present comparison test results for the calculation times for constructing the A-matrix and for solving the linear equations to obtain the dipole moments.

Note that the full equivalent number of dipoles are plotted (figures \ref{fig:calc_A} \& \ref{fig:calc_P}) for the case of the rotational symmetric algorithm and not just the number of dipoles in the quadrant.\\
Although the size of the 4-fold rotationally symmetric interaction matrices are $1/16$th that of the conventional interaction matrix, the times required to assemble are comparable (figure \ref{fig:calc_A}) and even slightly more for the former because the algorithms (\ref{eqn:A_offdiag}) \& (\ref{eqn:A_ondiag}) still requires each component of the rotational counterpart in each quadrant to be calculated, including the transformation matrices (\ref{eqn:sph2cart_mat} \& \ref{eqn:cart2sph_mat}) for changing coordinate systems back and forth. Moreover, the rotationally symmetric A-matrices are azimuthal mode, $m$, dependent and thus requires $2N_{max}+1$ number of compressed \emph{T}-matrices; this number can reduced by exploiting $\mathbf{A}(m)=\mathbf{A}(-m)$ for $m$ even. 

The important advantage is that the memory footprint of the 4-fold discrete rotational symmetry optimized interaction matrix is $1/16$th that of its conventional counterpart. For the full interaction matrix, we encountered the problem of disk swapping on a PC with 4Gb of RAM just with $>5000$ dipoles and as can be seen on figure \ref{fig:calc_A}, we stopped short at $<5000$ dipoles because the time taken increased sharply as available memory became scarce. By contrast, the symmetry optimized interaction matrix did not encounter such problems for the range of number of dipoles we tested with.

The reduction ($1/16$th) of the number of linear equations when calculating the dipole moments results very significant time savings (figure \ref{fig:calc_P}). The comparison of the respective scaling with the number dipoles can be determined from the gradients of the log-log plot in figure \ref{fig:calc_P}b. 
\subsection{Phase functions for the spheres and cubes}
To test the integrity of the optimization methods and the T-matrix via point matching, we compare against the results of the GLMT functions for the sphere~\cite{vandeHulst1957} and cube~\cite{Wriedt1998a}. The phase functions in figure \ref{fig:NF_sphere_phasefn} were generated from the scatterring coefficients obtained from the T-matrices; the results show good agreement against the Mie solution, allowing for stair casing errors when trying to approximate a sphere with a cubic lattice arrangement of dipoles with finite lattice spacing. 

The rotational symmetry algorithm was tested for cubes, since they possess 4-fold rotational symmetry and analytical results can be calculated. The T-matrix via point matching was also applied and the results (figure \ref{fig:WC_cube}) compared favourably against those produced from the GLMT method in \cite{Wriedt1998a}.
\subsection{\label{xrotor}Cross rotor torque calculations}
All the optimization methods and the point matching method for formulating the T-matrix were applied to calculating the forces and torques imparted on the microrotor (figure \ref{fig:cross}a) by a tightly focuused and trapping $LG_{02}$ laser beam. We were able to determine the axial equilibrium position (figure \ref{fig:force_torque}a) i.e. the point at which the axial force curve crosses the z-axis with negative gradient. The torque varies dramatically as we move along the beam axis but we are only interested in the torque at equilibrium. There are a number of uncertainties in experimental measurement that we endeavour to narrow down but the results between the model and experiment agree within the error limits.
\section{Discussion}
The rotational and mirror symmetry algorithms do not introduce any errors (well below round-off errors) to the DDA calculation. The construction of the interaction matrix takes a large portion of the calculation time for the DDA method in general; in principle, the compressed interaction matrix offers no advantage here. Because the compressed interaction matrix is azimuthal mode dependent, it can be cumbersome when formulating the T-matrix; we load the precalculated matrices as required. However, there are two major advantages of the symmetry optimization scheme. Firstly, their memory footprint is significantly smaller; we are otherwise unable to model our microrrotors, given their size, on desktop PCs. Secondly, the time savings when calculating the dipole moments more than make up for the extra time spent for the compressed interaction matrices.

The mode redundancy algorithms give further time savings of about two orders of magnitude, as with the mode-reduced T-matrices where applicable. 

There are a number of factors that contribute to errors and uncertainties between the quantities of interest between experimental and modelling results. In our DDA model, we assume that the material is linear and non-absorbing. The lattice spacing is finite, which leads to stair casing errors and an inaccurate representation of the target. The characterics of the driving beam does invariably differ to some degree compare to that of the intended beam due to imperfections in the optical setup. Taking all these factors into consideration, the methods in described in this paper gives us sufficient accuracy in modelling light scattering from the micromachines that we continue to design and develop.

\pagebreak
\bibliographystyle{elsarticle-num}



%
\end{document}